\documentclass[%preprint,%
twocolumn,%
amsmath,floatfix,nofootinbib,aps,prd]{revtex4}

\usepackage{graphicx}% Include figure files
\usepackage{bm}% bold math
\usepackage{axodraw}

\allowdisplaybreaks

\newcommand{\extranewlineforpreprintmode}{}
\newcommand{\MSbar}{{\overline{\rm MS}}}
\newcommand{\DRbar}{{\overline{\rm DR}}}
\newcommand{\DRbarprime}{\overline{\rm DR}'}
\newcommand{\deltaMSbar}{\delta_{\overline{\rm MS}}}
\newcommand{\lnbar}{{\overline{\rm ln}}}
\newcommand{\Fbar}{\overline{F}}
\newcommand{\Tbar}{\overline{T}}
\newcommand{\Vbar}{\overline{V}}
\newcommand{\propA}{{\rm A}}
\newcommand{\propB}{{\rm B}}
\newcommand{\propS}{{\rm S}}
\newcommand{\propU}{{\rm U}}
\newcommand{\propM}{{\rm M}}
\newcommand{\propV}{{\rm V}}
\newcommand{\propW}{{\rm W}}
\newcommand{\propX}{{\rm X}}
\newcommand{\propY}{{\rm Y}}
\newcommand{\propZ}{{\rm Z}}
\newcommand{\propgaugeSS}{G_{SS}}
\newcommand{\propgaugeFF}{G_{FF}}
\newcommand{\propgaugeFbarFbar}{G_{\Fbar\Fbar}}
\newcommand{\propgaugeSSSS}{G_{SSSS}}
\newcommand{\propgaugeSSFF}{G_{SSFF}}
\newcommand{\propgaugeSSFbarFbar}{G_{SS\Fbar\Fbar}}
\newcommand{\re}{\text{Re}}
\renewcommand{\theequation}{\arabic{section}.\arabic{equation}}
\begin{document}

hep-ph/0312092 
%% Unfortunately, revtex4 requires a silly line like the previous one, 
%% otherwise it later will barf on axodraw figures in twocolumn mode. 
%% Don't ask me why!

\title{Two-loop scalar self-energies in a general renormalizable theory\\
at leading order in gauge couplings}

\author{Stephen P. Martin}

\affiliation{
Physics Department, Northern Illinois University, DeKalb IL 60115 USA\\
{\rm and}
Fermi National Accelerator Laboratory, PO Box 500, Batavia IL 60510}

\begin{abstract} 
I present results for the two-loop self-energy functions for scalars in a
general renormalizable field theory, using mass-independent
renormalization schemes based on dimensional regularization and
dimensional reduction.  The results are given in terms of a minimal set of
loop-integral basis functions, which are readily evaluated numerically by
computers.  This paper contains the contributions corresponding to the
Feynman diagrams with zero or one vector propagator lines. These are the
ones needed to obtain the pole masses of the neutral and charged Higgs
scalar bosons in supersymmetry, neglecting only the purely electroweak
parts at two-loop order. A subsequent paper will present the results for
the remaining diagrams, which involve two or more vector lines.
\end{abstract}

\maketitle

%% Unfortunately, the combination of twocolumn mode and \tableofcontents
%% seems to give bad results at the arXiv. I don't know why. 
%% So, for now, put \tableofcontents in single column format, even
%% though it is uglier.
\onecolumngrid
\tableofcontents
\vspace{0.2in}\twocolumngrid

\protect\section{\label{sec:introduction}\protect{Introduction}}
\setcounter{equation}{0}

The next generation of high-energy collider experiments, including the
Large Hadron Collider (LHC) and a future electron-positron linear collider
(LC), should discover the mechanism of electroweak symmetry breaking. They
also hold the promise of very accurate measurements of the relevant
parameters of that mechanism. If supersymmetry is correct, then the
lightest neutral Higgs scalar boson mass should be measurable to an
accuracy of order 100 MeV at the LHC and even better at an LC
\cite{Higgsexp}. The mass of the top quark, which is known to be strongly
coupled to the electroweak symmetry breaking sector, can be obtained with
an accuracy of order 1 GeV at the LHC \cite{Beneke:2000hk} and 100 MeV at
an LC \cite{Abe:2001np}. These will easily exceed the accuracy obtained by
theoretical calculations at one-loop order in the Standard Model and its
extensions.  Therefore, calculation at two-loop order, at least, will be
necessary for detailed comparisons of candidate models with experiment. In
minimal supersymmetry, for example, the higher-order corrections to the
effective potential and to the mass of the lightest neutral Higgs scalar
boson are especially important for two reasons.  First, effects
proportional to the relatively large QCD coupling first appear at two-loop
order.  Second, the tree-level scalar potential has a nearly flat
direction, so that loop corrections involving larger mass scales are
relatively enhanced.

A great deal of progress has already been made on calculations of two-loop
corrections to masses and self-energy functions, including specific
results for the top quark \cite{topmassstart}-\cite{topmassend} and the
electroweak gauge bosons in the Standard Model \cite{Jegerlehner:2001fb},
the Higgs scalar bosons in the Standard Model \cite{DG94} and 
supersymmetry
\cite{Hempfling:1994qq}-\cite{Martin:2002iu}, and methods for evaluating
two-loop self-energy integrals in general
\cite{generalstart}-\cite{Martin:2003qz}.

Calculations of two-loop corrections can be difficult.  Since we do not
yet know with certainty what the correct model of physics at the TeV scale
is, or even a complete list of its degrees of freedom, a prudently
flexible strategy is called for. Accordingly, I propose to compute
self-energy functions for particles in a completely general renormalizable
theory. These ``once-and-for-all'' results can then be specialized to any
perturbative model, as future developments warrant.\footnote{Of course,
the correct model of physics near the TeV scale might be
non-renormalizable, or non-perturbative, or not even described by a
four-dimensional field theory. But supersymmetry, at least, provides
grounds for optimism.} A similar approach was used for the effective
potential in ref.~\cite{Martin:2001vx}; I will use exactly the same
conventions and notations in the present paper.

I choose to use a mass-independent renormalization scheme based on
dimensional continuation with minimal subtraction, in which physically
measurable quantities (physical masses, cross-sections, decay rates, etc.)  
are always outputs, and the input parameters are renormalized running
masses and couplings.  One can then conduct a simultaneous global fit to
all observables. This is in contrast to on-shell schemes, which use some
preferred observables as inputs. The use of a mass-independent scheme
means that one does not need to commit in advance to a particular choice
of input observables. Another possible advantage of mass-independent
schemes is that the running input parameters can be evolved to higher mass
scales (for example, the Planck scale, the scale of unification of gauge
couplings, the supersymmetry-breaking scale, etc.)  using the
renormalization group, to compare with the predictions of various
candidate models. For non-supersymmetric models, the traditional $\MSbar$
scheme \cite{msbar} is used, while for models based on softly broken
supersymmetry one may use the $\DRbarprime$ scheme \cite{drbarprime},
which is based on dimensional reduction \cite{drbar} but eliminates the
appearance of unphysical epsilon scalar masses in the formulas.

Consider a general renormalizable theory containing vector bosons,
fermions, and scalar bosons, the latter labeled by indices $i,j,k,\ldots$.
The main objective of this paper is to compute the two-loop scalar 
self-energy
\begin{eqnarray}
\Pi_{ij}(s) = 
\frac{1}{16\pi^2} \Pi_{ij}^{(1)} +
\frac{1}{(16\pi^2)^2} \Pi_{ij}^{(2)} + \ldots
\end{eqnarray}
as a function of 
\begin{eqnarray}
s = -p^2,
\label{eq:defines}
\end{eqnarray}
where $p^\mu$ is the external momentum, and the metric is either of
signature ($-$$+$$+$$+$) or Euclidean. Note that $s$ is taken to be real
with an infinitesimal positive imaginary part to resolve the branch cuts.
The self-energy function is gauge-dependent, but can be used to obtain a
gauge-invariant \cite{Willenbrock:1991hu}-\cite{Gambino:1999ai} pole
squared mass, defined as the position of the complex pole, with
non-positive imaginary part, in the propagator obtained from the
perturbative Taylor expansion of the self-energy function about the
tree-level squared mass \cite{poleambiguity}. Explicitly, for a scalar
particle with tree-level renormalized (running) squared mass $m_k^2$, the
two-loop pole squared mass
\begin{eqnarray}
s_{k} = M_k^2 - i \Gamma_k M_k
\end{eqnarray} 
can be obtained iteratively as follows. First, the one-loop approximation
to the pole squared mass, $s_k^{(1)}$, is the appropriate\footnote{In this
discussion, the ``appropriate'' solutions are the ones that continuously 
go
over to $m_k^2$ as the loop corrections to the self-energy are formally
turned off by reducing the three- and four-particle couplings of the
theory to zero.} solution of
\begin{eqnarray}
\mbox{Det}\left [ (m_i^2- s_{k}^{(1)})
\delta_{ij} +  \frac{1}{16 \pi^2}\Pi^{(1)}_{ij}(m_k^2) \right ] = 0.
\label{eq:genonelooppole}
\end{eqnarray}
Then, defining
\begin{eqnarray}
[ \widetilde  \Pi_k ]_{ij} &=&
\frac{1}{16 \pi^2}  \left [\Pi^{(1)}_{ij}(m_k^2)
+ (s_{k}^{(1)} - m_k^2) \Pi^{(1)\prime}_{ij}(m_k^2) \right ]
\nonumber \\ &&
+ \frac{1}{(16 \pi^2)^2} 
\Pi^{(2)}_{ij}(m_k^2),
\phantom{xxx}
\label{eq:definepitilde}
\end{eqnarray}
the two-loop approximation to the pole squared mass $s_k^{(2)}$ is the 
appropriate 
solution to
\begin{eqnarray}
\mbox{Det}\left [ (m_i^2- s_k^{(2)})
\delta_{ij} +[ \widetilde  \Pi_k ]_{ij} \right ] = 0
.
\label{eq:gentwolooppole}
\end{eqnarray}
Repeated indices are not summed in the last four equations. The prime in
equation (\ref{eq:definepitilde}) indicates a derivative with respect to
the argument $s$, which is then replaced by $m_k^2$. Note that $[
\widetilde \Pi_k ]_{ij}$ is not a function of $s$.

In this paper, the contributions from diagrams involving zero or one
vector lines are included; the rest will appear in a subsequent paper.  
Note that these diagrams are already enough to compute the full two-loop
pole masses of all of the neutral and charged Higgs scalar bosons in
supersymmetric extensions of the Standard Model, neglecting only the
purely electroweak contributions (those that involve $g^4$, $g^2 g^{\prime
2}$ or $g^{\prime 4}$). The explicit results for the particular case of
the Minimal Supersymmetric Standard Model will also be presented in a
separate paper.

The rest of this paper is organized as follows. Section
\ref{sec:conventions} specifies the notations, conventions and methods
used. Section \ref{sec:oneloop} reviews the self-energy functions at one
loop order. Section \ref{sec:twoloop} reports the results for two-loop
scalar self-energies in the $\MSbar$ and $\DRbarprime$ schemes, with
vector bosons assumed to have a generic non-zero mass. In section
\ref{sec:massless}, the corresponding results for massless vectors are
obtained.  Section \ref{sec:examples} contains an illustrative example,
featuring several non-trivial consistency checks.

\protect\section{\label{sec:conventions}\protect{Notations and Setup}}
\setcounter{equation}{0}

In the following, I consider a general renormalizable field theory,
consisting of real scalars $R_i$, two-component Weyl fermions $\psi_I$,
and vector fields $A^\mu_a$. After expanding around the vacuum expectation
values of the scalars, the tree-level squared masses of these fields are
taken to be diagonalized, and are denoted by $m_i^2$, $m_I^2$, and $m_a^2$
respectively. This does not mean that the masses of the fermions are
necessarily diagonal, however. The fermion mass matrices $M^{IJ}$ and
$M_{IJ} \equiv (M^{IJ})^*$ appear in the numerators of chirality-changing
fermion propagators. They can be non-zero when the fermions $\psi_I$ and
$\psi_J$ have the same squared mass. In this squared-mass eigenstate
basis, the interaction Lagrangian terms have the form:
\begin{eqnarray}
{\cal L}_{\rm S} &=& - \frac{1}{6} \lambda^{ijk} R_i R_j R_k
- \frac{1}{24} \lambda^{ijkl} R_i R_j R_k R_l
,  \label{LS}
\\
{\cal L}_{\rm SF} &=& - \frac{1}{2} y^{IJk} \psi_I \psi_J R_k + {\rm
c.c.}
,
\\
{\cal L}_{\rm SV} &=& 
- g^{aij} A^\mu_a R_i \partial_\mu R_j
-\frac{1}{4} g^{abij}  A^\mu_a A_{\mu b} R_i R_j
\nonumber \\ && 
-\frac{1}{2} g^{abi} A^\mu_a A_{\mu b} R_i
,
\label{LSV}
\\
{\cal L}_{\rm FV} &=& g^{aJ}_I A_a^\mu \psi^{\dagger I} {\overline
\sigma}_\mu \psi_J
,
\label{LFV}
\end{eqnarray}
with repeated indices summed over, and metric signature ($-$$+$$+$$+$).
The squared masses $m^2_i$, $m^2_I$, $m^2_a$, the fermion 
masses $M^{IJ}$, and the couplings $\lambda^{ijk}$, 
$\lambda^{ijkl}$, $y^{IJk}$, $g^{aij}$,
$g^{abij}$, $g^{abi}$, and $g^{aJ}_I$ are all renormalized running 
parameters. Raising or lowering of fermion indices implies complex 
conjugation of the coupling, so:
\begin{eqnarray}
y_{IJk} \equiv (y^{IJk})^*, \qquad g^{aJ}_I \equiv (g^{aI}_J)^* .
\end{eqnarray}
(I have omitted possible three-vector, four-vector, and ghost
interactions, which will not play a role in this paper.) For more
discussion, see ref.~\cite{Martin:2001vx}.

The propagators for the scalars $R_i$ are diagonal, and are given 
for four-momentum $k^\mu$ by
\begin{eqnarray}
\frac{-i}{k^2 + m_i^2}.
\end{eqnarray}
The fermions have diagonal chirality-changing propagators,
\begin{eqnarray}
\frac{-i k \cdot \sigma}{k^2 + m_I^2},
\qquad \frac{i k \cdot \overline\sigma}{k^2 + m_I^2},
\end{eqnarray}
as well as chirality-violating propagators that are not, in general, 
diagonal:
\begin{eqnarray}
\frac{-i M^{IJ}}{k^2 + m_I^2},\qquad
\frac{-i M_{IJ}}{k^2 + m_I^2} .
\end{eqnarray}
In a general $R_\xi$ gauge, the
propagator for a vector boson with squared mass $m_a^2$  and four-momentum 
$k^\mu$ is
\begin{eqnarray}
\frac{i\eta^{\mu\nu}}{k^2 + m_a^2} + i {\cal L}_{m^2_a} 
\left (
\frac{k^\mu k^\nu}{k^2 + m_a^2} \right ),
\label{eq:vectorprop}
\end{eqnarray}
where I have introduced the notation
\begin{eqnarray}
{\mathcal L}_x f(x) \equiv \left [ f(x) - f( \xi x) \right ]/x ,
\end{eqnarray}
with $\xi=0$ for Landau gauge and $\xi=1$ for Feynman gauge.
Results below for loop-integral functions involving massive vector 
bosons will be conveniently
written in terms of the ${\mathcal L}_x$ notation.
For massless vector gauge fields, one has:
\begin{eqnarray}
\lim_{x \rightarrow 0} [ {\cal L}_x f(x)] = (1-\xi) f'(0) ,
\end{eqnarray}
corresponding to a propagator 
\begin{eqnarray}
i \eta^{\mu\nu}/k^2 + i (\xi-1) k^\mu k^\nu/(k^2)^2 .
\end{eqnarray}
Infrared divergences associated with massless vector bosons are dealt with
by first computing with a finite mass, and then taking the massless 
vector limit.

In preparing the Lagrangian as above, there is a choice to be made
regarding the vacuum expectation values of scalar fields. One procedure is
to expand around the VEVs that minimize the tree-level potential. This
would mean that there are no tree-level tadpole couplings. However, in
that case, there can be non-vanishing tadpole graphs at one-loop order and
beyond, which must be included in the self-energy functions.  A different
choice is to instead expand around VEVs that minimize the full
loop-corrected effective potential.  This means that the tree-level
potential, in general, will have tadpole coupling terms:
\begin{eqnarray}
{\cal L}_{\rm tadpole} = -\lambda^i R_i .
\end{eqnarray}
However, expanding around a minimum of the full loop-corrected effective
potential means that the sum of all tadpole contributions always vanishes,
through whatever loop order we are working. Therefore, as long as we
expand around a minimum of the full effective potential, we can safely
neglect all tadpole graphs, and the tree-level tadpole coupling
$\lambda^i$ will never appear explicitly in results; its only effect is to
precisely cancel the sum of the loop tadpoles, which therefore do not need
to be calculated either.

There is a subtlety in the case of spontaneous breaking of gauge
symmetries, however. The first method above applies for arbitrary $\xi$,
and independence of the pole masses with respect to $\xi$ provides a
valuable check. But, if we choose the second method, expanding around a
VEV that is not the minimum of the tree-level potential, then the results
obtained below without including tadpole graphs are only valid in Landau
gauge, $\xi=0$. This is because for $\xi \not=0$, there is a non-trivial
tree-level mixing between the longitudinal components of the massive gauge
field and the Goldstone bosons (found among the $R_i$), so that the
propagator does not have the simple form of equation
(\ref{eq:vectorprop}). This is no great loss in practice, since the
general two-loop corrections to the effective potential are only simple
and have only been evaluated in Landau gauge, anyway.

In this paper, I will not include tadpole loop graphs, as I favor the
method of expanding around the VEVs that minimize the full effective
potential, rather than VEVs that minimize only the tree-level potential.
This means that Landau gauge $\xi=0$ should be used, whenever it makes a
difference, in practical applications of this work involving massive
vector bosons.\footnote{However, the calculations of the individual graphs
do not actually use any particular minimization condition for the VEVs. So
one could also use the results given here, together with a further
calculation of the relevant tadpole graphs, to obtain the answers valid
for an expansion about the VEVs that minimize the tree-level potential.}

The one-loop and two-loop contributions to the self-energy function
are calculated below as the sum of connected, one-particle irreducible
Feynman graphs with two external scalar lines. 
The topologies for one-loop and two-loop self-energy graphs are shown in
Figures 1 and 2. 
\begin{figure}[t]
\begin{picture}(100,65)(-50,-28)
\SetWidth{0.9}
\Line(-40,-5)(0,-5)
\Line(40,-5)(0,-5)
\CArc(0,13)(18,0,360)
\Text(0,-25)[]{$\propA$}
\end{picture}
\begin{picture}(100,65)(-50,-28)
\SetWidth{0.9}
\Line(-40,3)(-20,3)
\Line(40,3)(20,3)
\CArc(0,3)(20,0,360)
\Text(0,-25)[]{$\propB$}
\end{picture}
\caption{\label{fig:onelooptopologies}One-loop self-energy topologies.}
\end{figure}
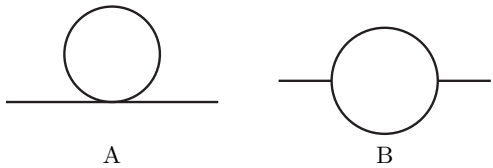
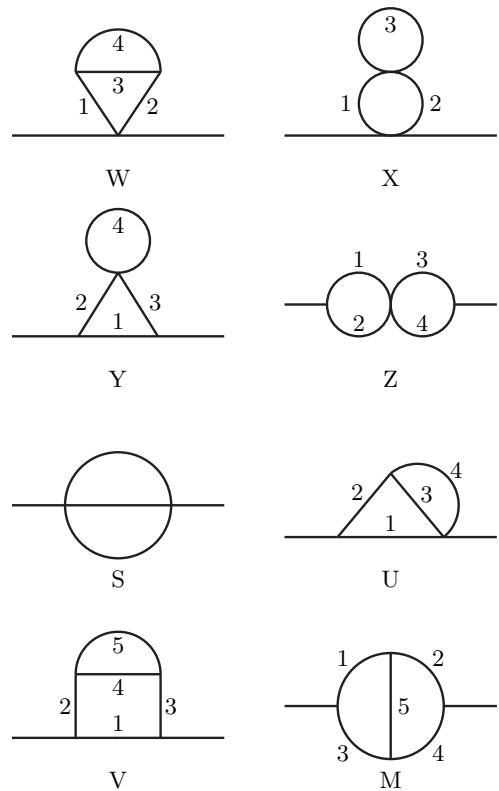
\begin{figure}[tb]
\begin{picture}(100,75)(-50,-35)
\SetWidth{0.9}
\Line(-40,-12)(0,-12)
\Line(40,-12)(0,-12)
\CArc(0,12)(16,0,180)
\Line(0,-12)(-16,12)
\Line(0,-12)(16,12)
\Line(-16,12)(16,12)
\Text(-13,-1)[]{$1$}
\Text(13,-1)[]{$2$}
\Text(0,7)[]{$3$}
\Text(0,23)[]{$4$}
\Text(0,-28)[]{$\propW$}
\end{picture}
\begin{picture}(100,75)(-50,-35)
\SetWidth{0.9}
\Line(-40,-12)(0,-12)
\Line(40,-12)(0,-12)
\CArc(0,0)(12,0,360)
\CArc(0,24)(12,0,360)
\Text(-17,0)[]{$1$}
\Text(17,0)[]{$2$}
\Text(0,30)[]{$3$}
\Text(0,-28)[]{$\propX$}
\end{picture}
\begin{picture}(100,75)(-50,-35)
\SetWidth{0.9}
\Line(-40,-12)(0,-12)
\Line(40,-12)(0,-12)
\CArc(0,24)(12,0,360)
\Line(-15,-12)(0,12)
\Line(15,-12)(0,12)
\Text(0,-6.5)[]{$1$}
\Text(-14,1)[]{$2$}
\Text(14,1)[]{$3$}
\Text(0,30)[]{$4$}
\Text(0,-28)[]{$\propY$}
\end{picture}
\begin{picture}(100,75)(-50,-35)
\SetWidth{0.9}
\Line(-40,0)(-24,0)
\Line(40,0)(24,0)
\CArc(-12,0)(12,0,360)
\CArc(12,0)(12,0,360)
\Text(-12,17)[]{$1$}
\Text(-12,-7)[]{$2$}
\Text(12,17)[]{$3$}
\Text(12,-7)[]{$4$}
\Text(0,-28)[]{$\propZ$}
\end{picture}
\begin{picture}(100,75)(-50,-35)
\SetWidth{0.9}
\Line(-40,0)(-20,0)
\Line(40,0)(20,0)
\CArc(0,0)(20,0,360)
\Line(-20,0)(20,0)
\Text(0,-28)[]{$\propS$}
\end{picture}
\begin{picture}(100,75)(-50,-35)
\SetWidth{0.9}
\Line(-40,-12)(-20,-12)
\Line(40,-12)(20,-12)
\Line(-20,-12)(0,12)
\Line(20,-12)(0,12)
\Line(-20,-12)(20,-12)
\CArc(10,0)(15.6205,-50.1944,129.806)
\Text(0,-7)[]{$1$}
\Text(-12.6,5)[]{$2$}
\Text(13.7,4)[]{$3$}
\Text(24.8,13.3)[]{$4$}
\Text(0,-28)[]{$\propU$}
\end{picture}
\begin{picture}(100,75)(-50,-35)
\SetWidth{0.9}
\Line(-40,-12)(0,-12)
\Line(40,-12)(0,-12)
\CArc(0,12)(16,0,180)
\Line(-16,-12)(-16,12)
\Line(16,-12)(16,12)
\Line(-16,12)(16,12)
\Text(0,-6.67)[]{$1$}
\Text(-20,0)[]{$2$}
\Text(20,0)[]{$3$}
\Text(0,7)[]{$4$}
\Text(0,23)[]{$5$}
\Text(0,-28)[]{$\propV$}
\end{picture}
\begin{picture}(100,75)(-50,-35)
\SetWidth{0.9}
\Line(-40,0)(-20,0)
\Line(40,0)(20,0)
\CArc(0,0)(20,0,360)
\Line(0,-20)(0,20)
\Text(-18,18)[]{$1$}
\Text(18,18)[]{$2$}
\Text(-18,-18)[]{$3$}
\Text(18,-18)[]{$4$}
\Text(5,0)[]{$5$}
\Text(0,-28)[]{$\propM$}
\end{picture}
\caption{\label{fig:twoloop}Two-loop self-energy topologies. 
The numbers correspond to a canonical 
ordering for the propagator lines, which applies both for
subscripts labeling propagator types (as $S$, $F$, $\Fbar$, or $V$) and 
for squared mass arguments in the corresponding loop integral functions.}
\end{figure}
The nomenclature used in this paper for the Feynman diagrams and their 
associated loop
functions is as follows. Letters A, B are assigned to the one-loop
topologies, and W, X, Y, Z, S, U, V, M for the two-loop topologies,
without regard to the types of propagators. For each such two-loop
topology, a canonical ordering is assigned to the internal propagator
lines, as shown. (No canonical ordering is needed for the one-loop
topologies or for the two-loop topology $\propS$.) Then the loop diagrams
are assigned a name given by the letter corresponding to the topology,
with subscripts $S$, $F$, $\Fbar$, or $V$ according to whether the
internal lines (in the canonical ordering) correspond to scalar,
chirality-preserving fermion, chirality-violating fermion, or vector
propagators. Diagrams related by interchange of the external scalar lines,
or by symmetries of the topology with respect to interchanges of internal
lines, are not considered separately. A one-loop example and a two-loop
example of this naming scheme are shown in Figure 3.
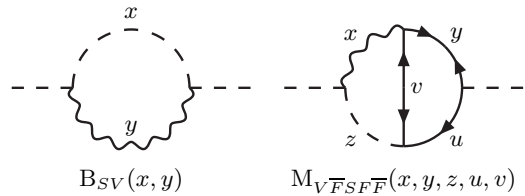
\begin{figure}[t]
\begin{picture}(100,75)(-50,-35)
\SetWidth{0.9}
\DashLine(-45,3)(-22,3){4.5}
\DashLine(45,3)(22,3){4.5}
\DashCArc(0,3)(22,0,180){5}
\PhotonArc(0,3)(22,180,360){1.6}{7.5}
\Text(0,31.5)[]{$x$}
\Text(0,-12.5)[]{$y$}
\Text(0,-32)[]{$\propB_{SV}(x,y)$}
\end{picture}
\begin{picture}(100,75)(-50,-35)
\SetWidth{0.9}
\DashLine(-45,3)(-22,3){4.5}
\DashLine(45,3)(22,3){4.5}
\PhotonArc(0,3)(22,90,180){1.6}{3.5}
\ArrowArc(0,3)(22,0,45)
\ArrowArcn(0,3)(22,90,45)
\ArrowArcn(0,3)(22,0,-90)
\DashCArc(0,3)(22,-180,-90){5}
\ArrowLine(0,3)(0,25)
\ArrowLine(0,3)(0,-19)
\Text(-20,23)[]{$x$}
\Text(20,23)[]{$y$}
\Text(-20,-17)[]{$z$}
\Text(20,-17)[]{$u$}
\Text(5,3)[]{$v$}
\Text(0,-32)[]{$\propM_{V\Fbar SF \Fbar}(x,y,z,u,v)$}
\end{picture}
\caption{\label{fig:twoloopex}Examples of the nomenclature for
self-energy diagram loop integral functions used in this paper.
Dashed lines represent scalars, wavy lines vectors, solid lines
with a single arrow are chirality-preserving fermion propagators,
and solid lines with clashing arrows are chirality-violating
fermion propagators. 
The arguments $x,y,z,u,v$ denote the corresponding
propagator squared masses.}
\end{figure}

For each Feynman diagram, 
the integrations over internal momenta are regulated by
continuing to $d = 4 - 2 \epsilon$ dimensions, according to
\begin{eqnarray}
\int d^4 k \rightarrow (2 \pi \mu)^{2 \epsilon} \int d^d k .
\end{eqnarray}
In the dimensional regularization scheme, the vector bosons also have $d$
components, while in the dimensional reduction scheme they have $d$
ordinary components and $2\epsilon$ additional components known as 
epsilon 
scalars, which in general have a different squared mass.
Counterterms for
the one-loop sub-divergences and the remaining two-loop divergences
are added, according to the rules of minimal subtraction, to give finite 
loop functions, which then depend on the renormalization scale $Q$ given 
by
\begin{eqnarray}
Q^2 = 4 \pi e^{-\gamma} \mu^2 .
\end{eqnarray}
The resulting renormalization schemes are known as $\MSbar$ \cite{msbar} 
and
$\DRbar$ \cite{drbar}, respectively. A further coupling constant 
redefinition in
the latter case removes the dependence on the unphysical epsilon scalar 
masses,
resulting in the $\DRbarprime$ scheme \cite{drbarprime} appropriate for 
(softly-broken)
supersymmetric models. 
In this paper, the symbol
\begin{eqnarray}
\deltaMSbar \equiv \begin{cases}
1 & \text{for}\quad\MSbar \\
0 & \text{for}\quad\DRbarprime
\end{cases}
\end{eqnarray}
will be used to present results simultaneously in the $\MSbar$ and 
$\DRbarprime$ schemes.

Below,
the result for each Feynman diagram is given in
terms of a corresponding loop integral function, for 
example
$\propB_{SV}(x,y)$ and $\propM_{V\Fbar S F \Fbar}(x,y,z,u,v)$ for the
Feynman diagrams shown in Figure 3. 
[These functions will be given explicitly in 
equations (\ref{eq:BSV}) and 
(\ref{eq:MVfSFf}) for massive vectors, 
and in equations
(\ref{eq:BSVzero}) and
(\ref{eq:MVfSFfzero}) for massless vectors.]
The variables $x,y,z,u,v$ here stand
for squared masses. Each loop function associated with a Feynman diagram
also depends on the external momentum invariant $s$, as defined in
equation (\ref{eq:defines}), and on the renormalization scale
$Q$, but these are not listed explicitly among the arguments, since they
are the same for all loop functions in a given calculation. By convention,
the usual symmetry factors associated with the Feynman diagram are
factored out of the corresponding loop function, but minus signs
associated with fermion loops and other factors associated with the
evaluation of the diagram are not. All counterterms are included 
within the loop integral functions.

Tarasov has described an algorithm
\cite{Tarasov:1997kx} for reducing any two-loop self-energy integral with
arbitrary masses to a minimal set of basis integral functions. This
algorithm has been conveniently implemented in the computer program TARCER
\cite{Mertig:1998vk}. In this paper, I use the Tarasov algorithm to reduce
the scalar two-loop self-energy integral functions to a (slightly 
modified) basis of functions that were defined in \cite{Martin:2003qz}. 

Briefly, the basis functions can be described as follows. At one loop,
integrals with scalar propagators and the topologies of $\propA$ and
$\propB$ in Figure 1 give rise to basis functions $A(x)$, $B(x,y)$, which
are essentially Passarino-Veltman \cite{Passarino:1978jh} functions. At
two loops, integrals with scalar propagators and the topologies $\propS$,
$\propU$, and $\propM$ in Figure 2 give rise to basis functions
$S(x,y,z)$, $U(x,y,z,u)$ and $M(x,y,z,u,v)$. Another basis function
$T(x,y,z)$ is obtained by differentiating $-S(x,y,z)$ with respect to $x$.
Also, $I(x,y,z)$ is defined as the two-loop vacuum integral function
obtained by setting $s=0$ in $S(x,y,z)$. These basis functions include
one-loop and two-loop counterterms, and they are finite and independent of
$\epsilon$. Their precise definitions were given in
ref.~\cite{Martin:2003qz}. That paper also gave a complete and efficient
algorithm for their numerical calculation, following the strategy put
forward in \cite{Caffo:1998du}-\cite{Caffo:2002wm} using the differential
equations method of \cite{Kotikov:1990kg}-\cite{Remiddi:1997ny}.

Because of the existence of graphs with the topology $\propV$ shown in
Figure 2 with the 2,3 propagators having the same squared mass, it is
useful as a matter of notational convenience to define the function
\begin{eqnarray}
V(x,y,z,u) \equiv -\frac{\partial}{\partial y} U(x,y,z,u).
\end{eqnarray}
This is not an independent basis function; instead, it can be 
expressed algebraically in terms of the basis 
functions, using equation (3.22) of \cite{Martin:2003qz}. 

It is useful to note that all of the functions mentioned in the previous
two paragraphs have smooth limits as squared-mass arguments are taken to
zero, with two exceptions; there are logarithmic singularities as
$x\rightarrow 0$ in the basis function $T(x,y,z)$ and as $y \rightarrow 0$
in $V(x,y,z,u)$. These infrared singularities must cancel from physical
quantities, but as a book-keeping device it is useful to define modified
versions of these functions that have smooth massless limits. Accordingly,
in reference \cite{Martin:2003qz}, the function
\begin{eqnarray}
\Tbar(0,y,z) \equiv \lim_{x \rightarrow 0} [T(x,y,z) + B(y,z) \lnbar x] 
\end{eqnarray}
was defined. Here we have used the notation that logarithms of 
dimensionful quantities are always expressed in terms of \begin{eqnarray}
\lnbar X \equiv \mbox{ln}(X/Q^2) .
\end{eqnarray}
An analytic expression for $\Tbar(0,y,z)$  was given in
equation (6.18) of \cite{Martin:2003qz}.
Similarly, it is useful to introduce 
\begin{eqnarray}
\Vbar (x,0,z,u) \equiv \lim_{y \rightarrow 0} \left [
V(x,y,z,u) - \frac{I(y',z,u)}{s-x}
\right ],
\label{eq:defVbar}
\end{eqnarray}
where the prime indicates a derivative with respect to that squared-mass
argument. The function $\Vbar (x,0,z,u)$ can be expressed algebraically in
terms of the basis functions; the result is in Appendix A of the present
paper.

Also, below I will use
\begin{eqnarray}
B(x',y) &\equiv& \frac{\partial}{\partial x} B(x,y) \\
I(x',y,z) &\equiv& \frac{\partial}{\partial x} I(x,y,z)
\end{eqnarray}
to express a few results more compactly than would otherwise be possible.
These derivatives, expressed algebraically in terms of the 
basis functions, are given in equations (3.1) and (5.3) of 
ref.~\cite{Martin:2003qz}; they are also easy to express
analytically in terms of logarithms and dilogarithms, using equations
(6.3)--(6.9) of that paper. Note that $B(x,y)$ and $I(x,y,z)$ are 
symmetric under interchange of arguments, so $B(x',y) = B(y,x')$ and
$I(x',y,z) = I(y,x',z)$ etc.

To summarize, here is the procedure for evaluating the two-loop scalar
self-energy function for a general renormalizable theory, given the
running parameters as inputs. The first step is to minimize the two-loop
effective potential \cite{Martin:2001vx} to find the VEVs of the theory.
Then, one expands the real scalar, two-component fermion, and real vector
fields about those VEVs, and diagonalizes the tree-level squared mass
matrices. This specifies the interaction Lagrangian in the form of
equations (\ref{LS})-(\ref{LFV}), defining the tree-level couplings.  
These can be plugged in to the formulas given in section III, IV, and V
below, which involve loop diagram functions [like $\propB_{SV}(x,y)$ and
$\propM_{V\Fbar SF \Fbar}(x,y,z,u,v)$], which are in turn written in terms
of basis functions:
\begin{eqnarray}
&&A(x),\>\> B(x,y),\>\> I(x,y,z),\>\> S(x,y,z),\>\> T(x,y,z), 
\nonumber \\ 
&&\Tbar(0,y,z),\>\> U(x,y,z,u),\>\> M(x,y,z,u,v),
\end{eqnarray}
and some additional functions
\begin{eqnarray}
B(x',y),\>\> I(x',y,z),\>\> V(x,y,z,u),\>\> \Vbar(x,0,z,u)\phantom{xx}
\end{eqnarray}
that can be written algebraically in terms of the basis
functions. Finally, the basis functions in general are evaluated by
computer, using the methods described in detail in \cite{Martin:2003qz}.

\protect\section{\label{sec:oneloop}\protect{One-loop scalar self-energy 
functions}}
\setcounter{equation}{0}

The one-loop self-energy function matrix for scalars in the general theory
specified in section \ref{sec:conventions} is:
\begin{widetext}
\begin{eqnarray}
\Pi^{(1)}_{ij} 
&=&
\frac{1}{2} \lambda^{ijkk} \propA_S (m_k^2) 
+ \frac{1}{2} \lambda^{ikl} \lambda^{jkl} \propB_{SS} (m_k^2,m_l^2) 
+ \re \bigl [ y^{KLi} y_{KLj} \bigr ] \propB_{FF} (m_K^2, m_L^2)  
\nonumber \\ &&
+ \re \bigl [ y^{KLi} y^{K'L'j} M_{KK'} M_{LL'} \bigr ]
   \propB_{\Fbar\Fbar} (m_K^2, m_L^2)
+ g^{aik} g^{ajk} \propB_{SV} (m_k^2, m_a^2)
+ \frac{1}{2} g^{aaij} \propA_{V} (m_a^2)
\nonumber \\ &&
+ \frac{1}{2} g^{abi} g^{abj} \propB_{VV} (m_a^2, m_b^2) ,
\end{eqnarray}
where repeated indices are summed over, and the loop integral functions 
are given 
by:
\end{widetext}
\begin{eqnarray}
\propA_{S}(x) &=&  A(x) ,
\\
\propB_{SS}(x,y) &=& - B(x,y) ,
\label{eq:defBSS}
\\
\propB_{FF}(x,y) &=&  (x+y-s) B(x,y) - A(x)-A(y),\phantom{xxx}
\label{eq:defBFF}
\\
\propB_{\Fbar\Fbar}(x,y) &=&  2 B(x,y) ,
\label{eq:defBFbarFbar}
\\
\propA_{V}(x) &=&  4A(x) + 2 x \deltaMSbar - {\mathcal L}_x [x A(x)] ,
\\
\propB_{SV}(x,y) &=& 
(2x-y+2s) B(x,y) + A(x) - 2 A(y) 
\nonumber \\ && 
\!\!\!\!\!\!\!\!\!\!\!\!
\!\!\!\!\!\!\!\!\!
+ {\mathcal L}_y [(x+y-s) A(y) - (x-s)^2 B(x,y)] ,
\label{eq:BSV}
\\
\propB_{VV}(x,y) &=&
-\frac{7}{2} B(x,y) 
+ 2 \deltaMSbar
+ \frac{1}{2} {\mathcal L}_x [x B(x,y)] 
\nonumber \\ &&
\!\!\!\!\!\!\!\!\!\!\!\!
\!\!\!\!\!\!\!\!\!
+ \frac{1}{2} {\mathcal L}_y [y B(x,y)]
+ \frac{1}{4} {\mathcal L}_x {\mathcal L}_y \bigl \lbrace
x A(y) + y A(x)
\nonumber \\ &&
\!\!\!\!\!\!\!\!\!\!\!\!
\!\!\!\!\!\!\!\!\!
+[ 2 s (x+y) -x^2 -y^2 -s^2 ] B(x,y) 
\bigr \rbrace  .
\end{eqnarray}
The last function $\propB_{VV}(x,y)$ combines the contributions
from the diagram with two vector propagators and the corresponding diagram
with two ghost propagators. 
In each case, the vector masses are taken to be non-zero. In the
limit of massless gauge vector bosons, one obtains:
\begin{eqnarray}
\propA_{V}(0) &=& 0 ,
\\
\propB_{SV}(x,0) &=& 
(3 - \xi) (x+s) B(0,x) 
\nonumber \\ &&
+ (3 - 2 \xi) A(x) 
+ 2 (\xi-1) s .
\label{eq:BSVzero}
\end{eqnarray}
As explained in section \ref{sec:conventions}, the $\MSbar$ scheme results 
are obtained by setting $\deltaMSbar=1$
in the above equations,
while the $\DRbarprime$ scheme results have $\deltaMSbar=0$.
Note that even though the function $\propB_{SV}(x,y)$ involves a vector 
propagator, it is the same in both
schemes, because the vector indices are contracted with momenta
rather than a metric tensor. The gauge-dependent parts
(acted on by an ${\cal L}_x$ or ${\cal L}_y$) do not differ between
the two schemes, for the same reason.

When computing a pole mass in the two-loop approximation, one needs
$
\Pi^{(1)\prime }(s) 
\equiv
\partial\Pi^{(1)}/\partial s.
$
This is easily obtained from the preceding, either by using the
analytical expression for the function $B(x,y)$, or by using the
identity:
\begin{eqnarray}
s\frac{\partial B(x,y)}{\partial s}  = 
-x B(x',y) -y B(x,y') -1,\phantom{x}
\label{eq:dBds}
\end{eqnarray}
which follows from equation (\ref{eq:dBdQ}) and dimensional analysis.

\protect\section{\label{sec:twoloop}\protect{Two-loop scalar self-energy 
functions}}
\setcounter{equation}{0}

In this section, I present the results for two-loop contributions to the
scalar self-energy. They are divided into subsections, depending on 
whether
the Feynman diagram contains internal propagators lines that are only 
scalars; scalars and fermions;
scalars and one vector; fermions and one vector; or
scalars, fermions, and one vector. Throughout the following, repeated
indices are summed over.

\protect\subsection{\label{subsec:S}\protect{Diagrams with only scalar 
propagators}}

The two-loop Feynman diagrams that involve only scalar propagator lines 
yield the following contribution to the self-energy: 
\begin{widetext}
\begin{eqnarray} 
&& \Pi^{(2)}_{ij} \>=\>
\frac{1}{4} \lambda^{ijkl} \lambda^{kmn} \lambda^{lmn}
\propW_{SSSS} (m_k^2, m_l^2, m_m^2, m_n^2)
+ \frac{1}{4} \lambda^{ijkl} \lambda^{klmm} \propX_{SSS}
(m_k^2, m_l^2, m_m^2)
\nonumber \\ && \quad
+ \frac{1}{2} \lambda^{ikl} \lambda^{jkm} \lambda^{lmnn} 
\propY_{SSSS} (m_k^2, m_l^2, m_m^2, m_n^2)
+ \frac{1}{4} \lambda^{ikl} \lambda^{jmn} \lambda^{klmn}
\propZ_{SSSS} (m_k^2, m_l^2, m_m^2, m_n^2)
\nonumber \\ && \quad
+ \frac{1}{6} \lambda^{iklm} \lambda^{jklm} \propS_{SSS} 
(m_k^2,m_l^2,m_m^2) 
+\frac{1}{2} \left (
\lambda^{ikl} \lambda^{jkmn} + \lambda^{jkl} \lambda^{ikmn} \right )
\lambda^{lmn} \propU_{SSSS}
(m_k^2, m_l^2, m_m^2, m_n^2) 
\nonumber \\ && \quad
+ \frac{1}{2} \lambda^{ikl} \lambda^{jkm} \lambda^{lnp} \lambda^{mnp} 
\propV_{SSSSS} (m_k^2, m_l^2, m_m^2, m_n^2, m_p^2)
\extranewlineforpreprintmode
+ \frac{1}{2} \lambda^{ikm} \lambda^{jln} \lambda^{klp} \lambda^{mnp}
\propM_{SSSSS}(m_k^2, m_l^2, m_m^2, m_n^2, m_p^2) .
\phantom{xxxx}
\end{eqnarray}
Here the loop integral functions 
are given by:
\begin{eqnarray}
\propW_{SSSS}(x,y,z,u) &=& [I(x,z,u) - I(y,z,u)]/(y-x) ,
\\
\propX_{SSS}(x,y,z) &=& A(z) [A(x)-A(y)]/(x-y) ,
\\
\propY_{SSSS}(x,y,z,u) &=& A(u) [B(x,z) - B(x,y)]/(y-z) ,
\\
\propZ_{SSSS}(x,y,z,u) &=& B(x,y) B(z,u) ,
\\
\propS_{SSS}(x,y,z) &=& -S(x,y,z) ,
\\
\propU_{SSSS}(x,y,z,u) &=& U(x,y,z,u) ,
\\
\propV_{SSSSS}(x,y,z,u,v) &=& [U(x, y, u, v) -U(x, z, u, v)]/(y - z) ,
\\
\propM_{SSSSS}(x,y,z,u,v) &=& -M(x,y,z,u,v) .
\end{eqnarray}
In several of these functions, denominators threaten to vanish when masses 
become degenerate. Taking the corresponding limits, one obtains:
\begin{eqnarray}
\propW_{SSSS}(x,x,z,u) &=& -I(x',z,u) ,
\\
\propX_{SSS}(x,x,z) &=& A(z)\, \lnbar x ,
\\
\propY_{SSSS}(x,y,y,u) &=& -A(u) B(x,y') ,
\\
\propV_{SSSSS}(x,y,y,u,v) &=& -V(x,y,u,v) .
\end{eqnarray}

\protect\subsection{\label{subsec:SF}\protect{Diagrams with scalar and 
fermion 
propagators}}

In this subsection, I present the results for Feynman diagrams that
involve both scalar and fermion propagators, but no vector propagators.

The contributions from diagrams with the topology $\propW$ in figure
\ref{fig:twoloop}, in which
propagators 1,2 are scalars and propagators 3,4 are fermions are:
\begin{eqnarray}
\Pi^{(2)}_{ij} &=&
\frac{1}{2} \lambda^{ijkl}
\re \bigl [y^{MNk} y^{M'N'l} M_{MM'} M_{NN'} \bigr ]
\propW_{SS\Fbar\Fbar} (m_k^2, m_l^2, m_M^2, m_N^2)
\nonumber \\ &&
+\frac{1}{2} \lambda^{ijkl} y^{MNk} y_{MNl}
\propW_{SSFF} (m_k^2, m_l^2, m_M^2, m_N^2),
\end{eqnarray}
where 
\begin{eqnarray}
\propW_{SS\Fbar\Fbar}(x,y,z,u) &=& -2 \propW_{SSSS}(x,y,z,u)
,
\\
\propW_{SSFF}(x,y,z,u) &=&
\left \lbrace (z+u-x) I(x,z,u)  -A(x) [A(z)+A(u)] \right \rbrace/(x-y)
+ (x \leftrightarrow y), \phantom{xxx}
\\
\propW_{SSFF}(x,x,z,u) &=&
(z+u-x) I(x',z,u) -I(x,z,u) -[A(u)+A(z)] \lnbar x .
\end{eqnarray}
The contributions from diagrams of the topology $\propM$, taking
propagators 1,2,3,4 to be fermions and 5 to be scalar, are:
\begin{eqnarray}
\Pi^{(2)}_{ij} &=&
\re \bigl [ y^{KMi} y^{LNj} y^{K'L'p} y^{M'N'p}
M_{KK'} M_{LL'} M_{MM'} M_{NN'} \bigr ]
\propM_{\Fbar\Fbar\Fbar\Fbar S}(m_K^2, m_L^2, m_M^2, m_N^2, m_p^2)
\nonumber \\ &&
+ 2 \re \bigl [ y^{KMi} y_{LNj} y_{KL'p} y^{M'Np} M^{LL'} M_{MM'} \bigr ]
\propM_{F\Fbar\Fbar FS}(m_K^2, m_L^2, m_M^2, m_N^2, m_p^2)
\nonumber \\ && 
+ \re \bigl [
\left ( y^{KMi} y_{LNj} + y^{KMj} y_{LNi} \right )
y_{KL'p} y_{MN'p} M^{LL'} M^{NN'} \bigr ]
\propM_{F\Fbar F\Fbar S}(m_K^2, m_L^2, m_M^2, m_N^2, m_p^2)
\nonumber \\ && 
+ 2 \re \bigl [ y^{KMi} y^{LNj} y_{KLp} y^{M'N'p} M_{MM'} M_{NN'} \bigr ]
\propM_{FF\Fbar\Fbar S}(m_K^2, m_L^2, m_M^2, m_N^2, m_p^2)
\nonumber \\ &&
+ \re \bigl [ y^{KMi} y^{LNj} y_{KLp} y_{MNp} \bigr ]
\propM_{FFFFS}(m_K^2, m_L^2, m_M^2, m_N^2, m_p^2) ,
\end{eqnarray}
where
\begin{eqnarray}
\propM_{\Fbar\Fbar\Fbar\Fbar S}(x,y,z,u,v) &=& 2 M(x,y,z,u,v) ,
\\
\propM_{F\Fbar \Fbar FS}(x,y,z,u,v) &=& 
(y+z-v-s) M(x,y,z,u,v) 
- U(x,z,u,v) 
\extranewlineforpreprintmode
- U(u,y,x,v) 
+ B(x,z) B(y,u) ,
\phantom{feif}
\\
\propM_{F\Fbar F\Fbar S}(x,y,z,u,v) &=& 
(x+z-s) M(x,y,z,u,v) - U(y,u,z,v) - U(u,y,x,v) ,
\\
\propM_{FF\Fbar\Fbar S}(x,y,z,u,v) &=& (x+y-v) M(x,y,z,u,v)
- U(x, z, u, v) - U(y, u, z, v) 
\extranewlineforpreprintmode
+ B(x,z) B(y, u)  ,
\\
\propM_{FFFFS}(x,y,z,u,v) &=&
(x u +y z- v s) M(x,y,z,u,v) 
-x U(z,x,y,v) -z U(x,z,u,v)
\nonumber \\ && 
-u U(y,u,z,v)
-y U(u,y,x,v)
+ S(x,u,v) + S(y,z,v)
\extranewlineforpreprintmode
+ s B(x,z) B(y,u) .
\end{eqnarray}
The results from diagrams of the topology $\propM$, taking
propagators 1,3 to be scalars and 2,4,5 to be fermions, are:
\begin{eqnarray}
\Pi^{(2)}_{ij} &=&
\lambda^{ikm} 
\Bigl (
\re \bigl [ y^{LNj} y^{L'Pk} y^{N'P'm} M_{LL'} M_{NN'} M_{PP'} \bigr ]
\propM_{S \Fbar S \Fbar \Fbar}(m_k^2, m_L^2, m_m^2, m_N^2, m_P^2)
\nonumber \\ && 
+
2 \re \bigl [ y^{LNj} y_{LPk} y^{N'Pm} M_{NN'} \bigr ]
\propM_{S F S \Fbar F}(m_k^2, m_L^2, m_m^2, m_N^2, m_P^2)
\nonumber \\ && 
+
\re \bigl [y^{LNj} y_{LPk} y_{NP'm} M^{PP'} \bigr ]
\propM_{S F S F \Fbar}(m_k^2, m_L^2, m_m^2, m_N^2, m_P^2)
\Bigr ) 
+ (i \leftrightarrow j) ,
\end{eqnarray}
where
\begin{eqnarray}
\propM_{S \Fbar S \Fbar \Fbar}(x,y,z,u,v) &=& 2 M(x,y,z,u,v) ,
\\
\propM_{S F S \Fbar F}(x,y,z,u,v) &=& (v - x +y) M(x,y,z,u,v)
+ U(y,u,z,v) - U(x,z,u,v) 
\extranewlineforpreprintmode
- B(x,z) B(y,u) 
,
\\
\propM_{S F S F \Fbar}(x,y,z,u,v) &=& 
(y+u-s) M(x,y,z,u,v) - U(x, z, u, v) - U(z, x, y, v) 
.
\end{eqnarray}

The contributions from diagrams of topology $\propV$, with lines
1,2,3 taken to be scalar and 4,5 to be fermions, are:
\begin{eqnarray} 
\Pi^{(2)}_{ij} &=&
\lambda^{ikl} \lambda^{jkm}
\Bigl ( 
\re \bigl [ y^{NPl} y^{N'P'm} M_{NN'} M_{PP'} \bigr ]
\propV_{SSS\Fbar\Fbar} (m_k^2, m_l^2, m_m^2, m_N^2, m_P^2)
\nonumber \\ && 
+ 
\re \bigl [y^{NPl} y_{NPm} \bigr ]
\propV_{SSSFF} (m_k^2, m_l^2, m_m^2, m_N^2, m_P^2)
\Bigr )
,
\end{eqnarray}
where
\begin{eqnarray}
\propV_{SSS\Fbar\Fbar}(x,y,z,u,v) &=&
-2 \propV_{SSSSS}(x,y,z,u,v) 
,
\\
\propV_{SSSFF}(x,y,z,u,v)
&=& 
\left \lbrace
(y-u-v) U(x,y,u,v) + [A(u)+A(v)]B(x,y) \right \rbrace /(y-z)
\extranewlineforpreprintmode
+ (y \leftrightarrow z)
,
\\
\propV_{SSSFF}(x,y,y,u,v)
&=&
(u+v-y) V(x,y,u,v) + U(x,y,u,v) +[A(u)+A(v)] B(x,y') 
.
\end{eqnarray}

The results from diagrams of topology $\propV$, with lines 1,2,3,4
taken to be fermions, and 5 to be scalar, are:
\begin{eqnarray}
\Pi^{(2)}_{ij} &=&
2 \re \bigl [
y^{KLi} y^{K'Mj} y^{L'Np} y^{M'N'p} M_{KK'} M_{LL'} M_{MM'} M_{NN'} 
\bigr ]
\propV_{\Fbar\Fbar\Fbar\Fbar S} (m_K^2, m_L^2, m_M^2, m_N^2, m_p^2) 
\nonumber \\ && 
+ 2 \re \bigl [ 
\bigl (y^{KLi} y^{K'Mj} +y^{KLj} y^{K'Mi} \bigr ) 
y_{LNp} y^{M'Np} M_{KK'} M_{MM'} 
\bigr ]
\propV_{\Fbar F\Fbar FS} (m_K^2, m_L^2, m_M^2, m_N^2, m_p^2)
\nonumber \\ &&
+ 2 \re \bigl [ y^{KLi} y^{K'Mj} y_{LNp} y_{MN'p} M_{KK'} M^{NN'} \bigr ]
\propV_{\Fbar FF \Fbar S} (m_K^2, m_L^2, m_M^2, m_N^2, m_p^2) 
\nonumber \\ &&
+ 2 \re \bigl [ y^{KLi} y_{KMj} y^{L'Np} y_{M'Np} M_{LL'} M^{MM'} \bigr ]
\propV_{F\Fbar\Fbar F S} (m_K^2, m_L^2, m_M^2, m_N^2, m_p^2)
\nonumber \\ &&
+ 2 \re \bigl [
\left ( y^{KLi} y_{KMj} +y^{KLj} y_{KMi} \right )
y_{LNp} y_{M'N'p} M^{MM'} M^{NN'} \bigr ]
\propV_{FF\Fbar\Fbar S} (m_K^2, m_L^2, m_M^2, m_N^2, m_p^2)
\nonumber \\ &&
+ 2 \re \bigl [
y^{KLi} y_{KMj} y_{LNp} y^{MNp} \bigr ]
\propV_{FFFFS} (m_K^2, m_L^2, m_M^2, m_N^2, m_p^2)
,
\end{eqnarray}
where
\begin{eqnarray}
\propV_{\Fbar\Fbar\Fbar\Fbar S}(x,y,z,u,v) &=& -2 \propV_{SSSSS}(x,y,z,u,v)
,
\\
\propV_{\Fbar F\Fbar FS}(x,y,z,u,v) &=&
\left \lbrace (v-y-u) U(x,y,u,v)
+[A(v)-A(u)] B(x,y) \right \rbrace/(y-z) 
\extranewlineforpreprintmode
+ (y \leftrightarrow z)
,
\\
\propV_{\Fbar FF\Fbar S}(x,y,z,u,v) &=&
2 [z U(x,z,u,v) - y U(x,y,u,v)]/(y-z)
,
\\
\propV_{F\Fbar\Fbar FS}(x,y,z,u,v) &=&
\bigl \lbrace
(y+u-v)[(s-x-y) U(x,y,u,v) + I(y,u,v)]
\nonumber \\ &&
+ [A(u) - A(v)] [(s-x-y) B(x,y) + A(y)]
\bigr \rbrace/2y(y-z) + (y \leftrightarrow z)
\nonumber \\ &&
+ \bigl \lbrace (u-v) [2x T(x,u,v) + 2 S(x,u,v) -A(x) -A(u) -A(v)
\extranewlineforpreprintmode
+x+u+v-s/4 ]
\nonumber \\ &&
+ u (x+u-v-s) T(u,v,x)
\extranewlineforpreprintmode
- v (x+v-u-s) T(v,x,u) \bigr \rbrace/2yz
,
\\
\propV_{FF\Fbar\Fbar S}(x,y,z,u,v) &=&
\bigl [ (s-x-y) U(x,y,u,v) + I(y,u,v) \bigr ]/(y-z) + (y \leftrightarrow z)
,
\\
\propV_{FFFFS}(x,y,z,u,v) &=& \bigl \lbrace
(y+u - v) [(s-x -y) U(x, y, u, v)+ I(y,u, v)] + y S(x,u,v)
\nonumber \\ &&
+ [A(u)-A(v)][ (s-x-y) B(x,y) + A(y) ]
\bigr \rbrace/2(y-z)
\extranewlineforpreprintmode
+ (y \leftrightarrow z)
.
\end{eqnarray}
In the case $y=z$, the necessary limits of these functions are given by:
\begin{eqnarray}
&&
\propV_{\Fbar F\Fbar FS}(x,y,y,u,v) =
(y+u-v) V(x,y,u,v) - U(x,y,u,v) 
\extranewlineforpreprintmode
+ [A(v)- A(u)] B(x,y')
,
\\
&&
\propV_{\Fbar FF\Fbar S}(x,y,y,u,v) = 2 y V(x,y,u,v) - 2 U(x,y,u,v)
,
\\
&&
\propV_{F\Fbar\Fbar FS}(x,y,y,u,v) =
\bigl \lbrace
y(y+u-v) [(x+y-s) V(x,y,u,v) + I(y',u,v)]
\extranewlineforpreprintmode
+[(u-v)(x-s) - y^2] U(x,y,u,v)
\nonumber \\ && \qquad\qquad
+ (u-v) [2 x T(x,u,v) +2S(x,u,v)-I(y,u,v) - A(x) -A(u)-A(v)
\extranewlineforpreprintmode
+x+u+v-s/4]
\nonumber \\ && \qquad\qquad
+ u(x+u-v-s) T(u,v,x) - v(x+v-u-s) T(v,x,u)
\nonumber \\ && \qquad\qquad
+ [A(u)-A(v)][(s-x-y) y B(x,y') + (x-s)B(x,y) +y] \bigr \rbrace/2y^2
,
\\
&&
\propV_{FF\Fbar\Fbar S}(x,y,y,u,v) =
(x+y-s) V(x,y,u,v) - U(x,y,u,v) + I(y',u,v)
\\
&&
\propV_{FFFFS}(x,y,y,u,v) =
\bigl \lbrace
(y+u-v) [(x+y-s) V(x,y,u,v) + I(y',u,v)]
\extranewlineforpreprintmode
+ (s - x - 2 y - u +v) U(x,y,u,v)
\nonumber \\ && \qquad\qquad
+S(x,u,v)
+I(y,u,v)
\extranewlineforpreprintmode
+ [A(u) - A(v)][(s-x-y)B(x,y') -B(x,y) + \lnbar y]
\bigr \rbrace/2.
\end{eqnarray}
The limit $y\rightarrow 0$ appropriate for massless fermions is
only needed when the corresponding
propagator has no mass insertions. For the case $
\propV_{\Fbar FF\Fbar S}(x,y,y,u,v)$, this limit is trivial,
since $y V(x,y,u,v)$ vanishes as $y\rightarrow 0$. The remaining
case is given by:
\begin{eqnarray}
\propV_{FFFFS}(x,0,0,u,v) &=&
\bigl \lbrace
(u-v)(x-s) \Vbar (x,0,u,v)
+ (s - x - u +v) U(x,0,u,v)
\extranewlineforpreprintmode
+S(x,u,v)
+I(0,u,v) \bigr \rbrace/2
\nonumber \\ &&
+ [A(u) - A(v)][s B(0,x') + A(x)/x]
.
\label{VFFFFSxoouv}
\end{eqnarray}
This function is well-defined as $s \rightarrow x$, 
although $(x-s) \Vbar (x,0,u,v)$ and $B(0,x')$ are both singular in that 
limit.

\protect\subsection{\label{subsec:SV}\protect{Diagrams
with scalar and one vector propagators}}

The contributions from two-loop Feynman diagrams with scalar propagators,
exactly one vector propagator, and no fermion propagators are
\begin{eqnarray}
&& \Pi^{(2)}_{ij} \>=\>
\frac{1}{2} \lambda^{ijkl} g^{akm} g^{alm}
\propW_{SSSV} (m^2_k, m^2_l, m^2_m, m^2_a )
+\frac{1}{4} \lambda^{ijkl} g^{aakl}
\propX_{SSV} (m^2_k, m^2_l, m^2_a )
\nonumber \\ &&
\qquad
+ \frac{1}{2} \lambda^{ikl} \lambda^{jkm} g^{aalm}
\propY_{SSSV} (m^2_k, m^2_l, m^2_m, m^2_a )
+ \frac{1}{2} g^{aik} g^{ajl} \lambda^{klmm}
\propY_{VSSS} (m^2_a, m^2_k, m^2_l, m^2_m)
\nonumber \\ &&
\qquad
+ \frac{1}{2}
( g^{aik} \lambda^{jkmn} + g^{ajk} \lambda^{ikmn} ) g^{amn}
\propU_{SVSS} (m^2_k, m^2_a, m^2_m, m^2_n )
\nonumber \\ &&
\qquad
+ \lambda^{ikl} \lambda^{jkm} g^{aln} g^{amn}
\propV_{SSSSV} (m^2_k, m^2_l, m^2_m, m^2_n, m^2_a )
\extranewlineforpreprintmode
+ \frac{1}{2} g^{aik} g^{ajl} \lambda^{kmn} \lambda^{lmn}
\propV_{VSSSS} (m^2_a, m^2_k, m^2_l, m^2_m, m^2_n)
\nonumber \\ &&
\qquad
+ \frac{1}{2} (g^{aik} \lambda^{jkl} + g^{ajk} \lambda^{ikl} )
g^{amn} \lambda^{lmn}
\propV_{SVSSS} (m^2_k, m^2_a, m^2_l, m^2_m, m^2_n)
\nonumber \\ &&
\qquad
+ \frac{1}{2} \lambda^{ikm} \lambda^{jln}
g^{akl} g^{amn} \propM_{SSSSV} (m^2_k, m^2_l, m^2_m, m^2_n, m^2_a )
\nonumber \\ &&
\qquad
+ ( g^{ail} \lambda^{jkm} + g^{ajl} \lambda^{ikm} ) g^{akn} \lambda^{lmn}
\propM_{VSSSS} (m^2_a,m^2_k, m^2_l, m^2_m, m^2_n)
,
\end{eqnarray}
where the loop functions are given by
\begin{eqnarray}
&& \propW_{SSSV}(x,y,z,u) \>=\>
\Bigl \lbrace (2x+2z -u) I(x,z,u)
+ A(x) [A(z) - 2 A(u)] + 2x A(u)
\nonumber \\ && \qquad\qquad
+ {\mathcal L}_u \left [ -(x-z)^2 I(x,z,u) + A(u) \lbrace
(z+u-x) A(x) + x A(z)\rbrace \right ] \Bigr \rbrace /(x-y)
\extranewlineforpreprintmode
+ (x \leftrightarrow y)
,
\\
&& \propW_{SSSV}(x,x,z,u) \>=\>
(2x+2z-u) I(x',z,u) +2 I(x,z,u)
\extranewlineforpreprintmode
+[A(z) - 2 A(u)] \, \lnbar x +2 A(u)
\nonumber \\ && \qquad
+ {\mathcal L}_u \bigl \lbrace -(x-z)^2 I(x',z,u) + 2 (z-x) I(x,z,u)
\extranewlineforpreprintmode
+ A(u) [ (z+u-2x) \lnbar x +x+A(z)] \bigr \rbrace
,
\\
&& \propX_{SSV}(x,y,z) \>=\>
\left (4 A(z)+ 2 z \deltaMSbar  - {\mathcal L}_z [z A(z)] \right
)[A(x)-A(y)]/(x-y)
,
\\
&& \propX_{SSV}(x,x,z) \>=\>
\left (4 A(z) + 2 z \deltaMSbar - {\mathcal L}_z [z A(z)] \right )\,
\lnbar x
,
\\
&& \propY_{SSSV}(x,y,z,u) \>=\>
\left (4 A(u) + 2 u \deltaMSbar - {\mathcal L}_u [u A(u)] \right )[B(x,z)-
B(x,y)]/(y-z)
,
\\
&& \propY_{SSSV}(x,y,y,u) \>=\>
-\left (4 A(u) + 2 u \deltaMSbar
- {\mathcal L}_u [u A(u)] \right ) B(x,y')
,
\\
&& \propY_{VSSS}(x,y,z,u) \>=\>  A(u) \bigl \lbrace
(2y-x+2s) B(x,y) + A(y)
\extranewlineforpreprintmode
+ {\mathcal L}_x \bigl [
y A(x) - (y-s)^2 B(x,y) \bigr ]
\bigr \rbrace/(y-z)
+(y \leftrightarrow z)
,
\\
&& \propY_{VSSS}(x,y,y,u) \>=\>  A(u) \bigl \lbrace
(2y-x+2s) B(x,y') + 2 B(x,y) + \lnbar y
\nonumber \\ && \qquad
+ {\mathcal L}_x \bigl [
A(x) - (y-s)^2 B(x,y') -2 (y-s) B(x,y) \bigr ]
\bigr \rbrace
,
\\
&& \propU_{SVSS}(x,y,z,u) \>=\> \bigl \lbrace
(z-u) \bigl [ (x-s) U(x,y,z,u) + 2 x T(x,z,u) + 2 S(x,z,u) 
\extranewlineforpreprintmode
- I(y,z,u)
-A(x) - A(z)
\nonumber \\ && \qquad
- A(u) +x+z+u-s/4 \bigr ]
+ z (x+z-u-s) T(z,u,x) 
\extranewlineforpreprintmode
-u (x+u-z-s) T(u,x,z)
\nonumber \\ && \qquad
+ [A(z) - A(u)] [(x-s) B(x,y) - A(y)]
\bigr \rbrace/y
\extranewlineforpreprintmode
+ {\mathcal L}_y \bigl \lbrace
(z-u) [(s-x) U(x,y,z,u)
+ I(y,z,u)]
\nonumber \\ && \qquad
+[A(z)-A(u)][(s-x) B(x,y) + A(y)] \bigr \rbrace
,
\\
&& \propV_{SSSSV}(x,y,z,u,v) \>=\> \Bigl \lbrace
(v-2y-2u) U(x,y,u,v) + [2 A(v) - A(u)] B(x,y)
\nonumber \\ && \qquad
+ {\mathcal L_v} \left [
(y-u)^2 U(x,y,u,v) -y S(x,u,v) + (y-u-v) A(v) B(x,y)
\right ] \Bigr \rbrace/(y-z)
\extranewlineforpreprintmode
+ (y \leftrightarrow z)
,
\\
&& \propV_{SSSSV}(x,y,y,u,v) \>=\>
(2y+2u-v) V(x,y,u,v) -2 U(x,y,u,v) 
\extranewlineforpreprintmode
+ [2 A(v) - A(u)] B(x,y')
\nonumber \\ && \qquad
+ {\mathcal L_v} \bigl \lbrace
2 (y-u) U(x,y,u,v)
-(y-u)^2 V(x,y,u,v)
-S(x,u,v)
\extranewlineforpreprintmode
+ A(v) [  (y-u-v) B(x,y') + B(x,y) ]
\bigr \rbrace
,
\\
&& \propV_{VSSSS}(x,y,z,u,v) \>=\>
\bigl \lbrace (x-2y-2s) U(x,y,u,v) - I(y,u,v)
\nonumber \\ && \qquad
+ {\mathcal L_x} \left [
(y-s)^2 U(x,y,u,v) - y S(x,u,v) \right ] \bigr \rbrace/(y-z)
+ (y \leftrightarrow z)
,
\\
&& \propV_{VSSSS}(x,y,y,u,v) \>=\>
(2y-x+2s) V(x,y,u,v) -2 U(x,y,u,v) - I(y',u,v)
\nonumber \\ && \qquad
+ {\mathcal L_x} \left [
2 (y-s) U(x,y,u,v) -(y-s)^2 V(x,y,u,v) - S(x,u,v) \right ]
,
\\
&& \propV_{SVSSS}(x,y,z,u,v) \>=\>
\Bigl \lbrace (u-v) [(x-s) U(x,y,u,v) -I(y,u,v)]
\extranewlineforpreprintmode
+ [A(u)-A(v)] [(x-s) B(x,y)
- A(y)] \Bigr \rbrace/y (y-z)
\nonumber \\ && \qquad
+ (y \leftrightarrow z)
+ \Bigl \lbrace (v-u) [2 x T(x,u,v) + 2 S(x,u,v)
- A(x) -A(u)
-A(v) 
\extranewlineforpreprintmode
+x+u+v-s/4 ]
\nonumber \\ && \qquad
+ u (s+v-x-u) T(u,x,v) + v (x+v-u-s) T(v,x,u) \Bigr \rbrace/yz
\extranewlineforpreprintmode
+ {\mathcal L}_y \Bigl [
\Bigl \lbrace
(u-v)[(s-x) U(x,y,u,v)
\nonumber \\ && \qquad
+ I(y,u,v)]
+ [A(u)-A(v)][(s-x) B(x,y) + A(y)]
\Bigr \rbrace/(y-z)
+ (y \leftrightarrow z) \Bigr ]
,
\\
&&
\propV^{\xi \not= 1}_{SVSSS}(x,y,y,u,v) \>=\>
\Bigl \lbrace (v-u) [
(x-s) U(x,y,u,v) - I(y,u,v)
+ 2 x T(x,u,v) 
\extranewlineforpreprintmode
+ 2 S(x,u,v)
- A(x) -A(u)
\nonumber \\ && \qquad
-A(v) +x+u+v-s/4 ]
+ u (s+v-x-u) T(u,x,v) + v (x+v-u-s) T(v,x,u)
\nonumber \\ && \qquad
+[A(u) - A(v)][(s-x) B(x,y) + A(y)]
\Bigr \rbrace/y^2
\extranewlineforpreprintmode
+ \frac{1}{(1-\xi) y}
{\cal L}_y \Bigl \lbrace (u-v) [(x-s) U(x,y,u,v)
-I(y,u,v)]
\nonumber \\ && \qquad
+ [A(u)-A(v)] [(x-s) B(x,y)
- A(y)] \Bigr \rbrace
,
\\
&&
\propV^{\xi=1}_{SVSSS}(x,y,y,u,v) \>=\>
\Bigl \lbrace (v-u) [
y (x-s) V(x,y,u,v) + (x-s) U(x,y,u,v)
\extranewlineforpreprintmode
+ y I(y',u,v) - I(y,u,v)
\nonumber \\ && \qquad
+ 2 x T(x,u,v) + 2 S(x,u,v)
- A(x)
-A(u)
-A(v) +x+u+v-s/4 ]
\extranewlineforpreprintmode
+ u (s+v-x-u) T(u,x,v)
\nonumber \\ && \qquad
+ v (x+v-u-s) T(v,x,u)
+ [A(u)-A(v)] [(x-s) y B(x,y') 
\extranewlineforpreprintmode
- (x-s) B(x,y) -y]
\Bigr \rbrace/y^2
,
\\
&&
\propM_{VSSSS}(x,y,z,u,v) \>=\>
(s-x+y+z-2u+v) M(x,y,z,u,v) 
\extranewlineforpreprintmode
- U(x,z,u,v)
+ 2 U(z,x,y,v)
\nonumber \\ && \qquad
-U(u,y,x,v)
+ U(y,u,z,v) - B(x,z) B(y,u)
+ {\mathcal L}_x \Bigl [
(s-z) [(v-y) M(x,y,z,u,v)
\nonumber \\ && \qquad
+ U(x,z,u,v) - B(x,z) B(y,u)]
+ (v-y) U(u,y,x,v) + S(x,u,v) 
\extranewlineforpreprintmode
-A(x) B(y,u) \Bigr ]
,
\\
&&
\propM_{SSSSV}(x,y,z,u,v) \>=\> \Bigl (
(x-v/4 -s/2) M(x,y,z,u,v) - U(x,z,u,v) + B(x,z) B(y,u)/4
\nonumber \\ && \qquad
+ {\mathcal L}_v \Bigl [
(y-x)[z M(x,y,z,u,v)/2 + U(z,x,y,v)] + S(x,u,v)/2  - A(v) B(x,z)/2 \Bigr ]
\Bigr )
\nonumber \\ && \qquad
+ (x \leftrightarrow y\>\mbox{and}\> z \leftrightarrow u)
+ (x \leftrightarrow z\>\mbox{and}\> y \leftrightarrow u)
+ (x \leftrightarrow u\>\mbox{and}\> y \leftrightarrow z)
.
\end{eqnarray}

\protect\subsection{\label{subsec:FV}\protect{Diagrams with fermion and 
one 
vector propagators}}

This subsection contains the contributions to the scalar self-energy
coming from two-loop diagrams
that involve fermion propagators, one vector propagator, and zero
scalar propagators.

The results for diagrams with topology $\propM$ with propagators
1,2,3,4 taken to be fermions and 5 to be vector are:
\begin{eqnarray}
&& \Pi^{(2)}_{ij} \>=\>
\re \bigl [ y^{KMi} y_{LNj} g^{aK'}_{L'} g^{a M'}_{N'}
M_{KK'} M^{LL'} M_{MM'} M^{NN'} \bigr ]
\propM_{\Fbar\Fbar\Fbar\Fbar V}(m_K^2, m_L^2, m_M^2, m_N^2, m_a^2)
\nonumber \\ &&
\qquad
+ 2 \re \bigl [ y^{KMi} y^{LNj} g^{a L'}_{K} g^{aM'}_{N} M_{LL'}  M_{MM'}
\bigr ]
\propM_{F\Fbar\Fbar F V}(m_K^2, m_L^2, m_M^2, m_N^2, m_a^2)
\nonumber \\ &&
\qquad
+ \re \bigl [
\bigl (y^{KMi} y^{LNj} + y^{KMj} y^{LNi} \bigr )
g^{aL'}_{K} g^{a N'}_{M} M_{LL'} M_{NN'} \bigr ]
\propM_{F \Fbar F \Fbar V}(m_K^2, m_L^2, m_M^2, m_N^2, m_a^2)
\nonumber \\ &&
\qquad
+ 2 \re \bigl [ y^{KMi} y_{LNj} g^{aL}_{K} g^{a M'}_{N'} M_{MM'} M^{NN'}
\bigr ] \propM_{FF \Fbar\Fbar V}(m_K^2, m_L^2, m_M^2, m_N^2, m_a^2)
\nonumber \\ &&
\qquad
+
\re \bigl [y^{KMi} y_{LNj} g^{aL}_{K} g^{a N}_{M} \bigr ]
\propM_{FFFFV}(m_K^2, m_L^2, m_M^2, m_N^2, m_a^2)
,
\end{eqnarray}
where
\begin{eqnarray}
&& \propM_{\Fbar \Fbar \Fbar \Fbar V}(x,y,z,u,v) \>=\>
8 M(x,y,z,u,v) - {\mathcal L}_v [2 v M(x,y,z,u,v)]
,
\\
&&\propM_{F \Fbar \Fbar F V}(x,y,z,u,v) \>=\>
\Bigl \lbrace (2y-v-s) M(x,y,z,u,v) - 2 U(x,z,u,v) 
\extranewlineforpreprintmode
+ B(x,z) B(y,u)
+ \deltaMSbar
\nonumber \\ && \qquad
+ {\mathcal L}_v \bigl [
\lbrace v(s/2-x) + (x-y)(z-u)/2\rbrace  M(x,y,z,u,v)
\extranewlineforpreprintmode
+(z-u) U(x,z,u,v) + (v+x-y) U(z,x,y,v)
\nonumber \\ && \qquad
-S(x,u,v)/2 - S(y,z,v)/2 + A(v) B(x,z) \bigr ] \Bigr \rbrace +
(x \leftrightarrow u \>\mbox{and}\> y \leftrightarrow z)
,
\\
&& \propM_{F \Fbar F \Fbar V}(x,y,z,u,v) \>=\> \Bigl \lbrace
(4x-2s)  M(x,y,z,u,v) - 4 U(y,u,z,v)
+ [2 B(y,u) + 1] \deltaMSbar
\nonumber \\ && \qquad
+ {\mathcal L}_v \bigl [v (s/2-x)  M(x,y,z,u,v) + v U(y,u,z,v)\bigr ]
\Bigr \rbrace
+ (x \leftrightarrow z\>\mbox{and}\> y \leftrightarrow u)
,
\\
&& \propM_{FF \Fbar \Fbar V}(x,y,z,u,v) \>=\> \Bigl [
(2x-v) M(x,y,z,u,v) - 2 U(x,z,u,v) + B(x,z) B(y,u)
+ \deltaMSbar
\nonumber \\ && \qquad
+ {\mathcal L}_v \Bigl \lbrace
2 z x (x-y-v) M(x,y,z,u,v)
\extranewlineforpreprintmode
+ [s (u-z-v) + (x+z) (v-u)
+ z (3x-2y+z)] U(x,z,u,v)
\nonumber \\ && \qquad
+ 2 x (v-u) T(x,u,v)
+ u (s+v-x-u) T(u,x,v)
+ v (x+v-u-s) T(v,x,u)
\nonumber \\ && \qquad
+ (2 v-2u-z) S(x,u,v)
+ (u-z-v) I(u,z,v)
\extranewlineforpreprintmode
+ A(v) [(x+z-s) B(x,z) - A(z) + u-v]
\nonumber \\ && \qquad
+ v [v+x-A(x)-A(u) -s/4] \Bigr \rbrace/2z \Bigr ]
+ (x \leftrightarrow y\>\mbox{and}\> z \leftrightarrow u)
,
\\
&&\propM_{FFFFV}(x,y,z,u,v) \>=\> \biggl (
(s-x-z)[ (s-y-u) M(x,y,z,u,v)/2 + 2 U(x,z,u,v)] 
\extranewlineforpreprintmode
+ S(x,u,v)
+ I(x,y,v)
\nonumber \\ && \qquad
+ [(x+z-s) B(x,z) - 2 A(x) + 2 x + v/2 -s/4] \deltaMSbar
\nonumber \\ && \qquad
+ {\mathcal L}_v \Bigl [
[ s z (x-y) + x (2 x u - 2 y z - v u) ] M(x,y,z,u,v)
+ [s (z-u-v) + (x+z)(u+v)
\nonumber \\ && \qquad
+ z (x-2y-z)] U(x,z,u,v)+ \lbrace x (s-x-3u+3v) T(x,u,v) + v (v-s) T(v,x,u)
\nonumber \\ && \qquad
+ (v-4x+3y-s) S(x,u,v)
+ A(v) [x-A(x)-v] 
\extranewlineforpreprintmode
+ v [v+x - 2 A(x) - 3 I(x,y,v) -s/4]
\rbrace/3 \Bigr ]/2 \biggr )
\nonumber \\ &&
\qquad
+ (x \leftrightarrow y\>\mbox{and}\> z \leftrightarrow u)
+ (x \leftrightarrow z\>\mbox{and}\> y \leftrightarrow u)
+ (x \leftrightarrow u\>\mbox{and}\> y \leftrightarrow z)
.
\end{eqnarray}

The results for diagrams of topology $\propV$ with lines 1,2,3,4 taken to
be fermion and 5 to be vector are:
\begin{eqnarray}
\Pi^{(2)}_{ij} &=&
2 \re \left [ y^{KLi} y^{K'Mj} g^{aL'}_{N} g^{a M'}_{N'}
M_{KK'} M_{LL'} M_{MM'} M^{NN'} \right ]
\propV_{\Fbar\Fbar\Fbar\Fbar V} (m_K^2, m_L^2, m_M^2, m_N^2, m_a^2)
\nonumber \\ &&
+2 \re \left [y^{KLi} y^{K'Mj} g^{aN}_{L} g^{a N'}_{M} M_{KK'} M_{NN'}
\right ]
\propV_{\Fbar FF\Fbar V} (m_K^2, m_L^2, m_M^2, m_N^2, m_a^2)
\nonumber \\ &&
+ 2 \re \left [ \left (y^{KLi} y_{KMj} + y^{KLj} y_{KMi} \right )
g^{aN}_{L} g^{a N'}_{M'} M^{MM'} M_{NN'} \right ]
\propV_{FF\Fbar\Fbar V} (m_K^2, m_L^2, m_M^2, m_N^2, m_a^2)
\nonumber \\ &&
+2 \re \left [ \left ( y^{KLi} y^{K'Mj} + y^{KLj} y^{K'Mi} \right )
g^{aN}_{L} g^{a M'}_{N} M_{KK'} M_{MM'} \right ]
\propV_{\Fbar F\Fbar F V} (m_K^2, m_L^2, m_M^2, m_N^2, m_a^2)
\nonumber \\ &&
+2 \re \left [ y^{KLi} y_{KMj} g^{aL'}_{N} g^{a N}_{M'} M_{LL'} M^{MM'}
\right ]
\propV_{F\Fbar\Fbar F V} (m_K^2, m_L^2, m_M^2, m_N^2, m_a^2)
\nonumber \\ &&
+2 \re \left [ y^{KLi} y_{KMj} g^{aN}_{L} g^{a M}_{N} \right ]
\propV_{FFFFV} (m_K^2, m_L^2, m_M^2, m_N^2, m_a^2)
,
\end{eqnarray}
where the integral functions are (including cases with degenerate
fermion masses $y=z$):
\begin{eqnarray}
&&
\propV_{\Fbar \Fbar \Fbar \Fbar V}(x,y,z,u,v) =
\left \lbrace -8 U(x,y,u,v) + 4 B(x,y) \deltaMSbar + {\mathcal L}_v [2v
U(x,y,u,v)]
\right \rbrace/(y-z) 
\extranewlineforpreprintmode
+ (y \leftrightarrow z)
,
\\
&&
\propV_{\Fbar \Fbar \Fbar \Fbar V}(x,y,y,u,v) =
8 V(x,y,u,v)  + 4 B(x,y') \deltaMSbar - {\mathcal L}_v [2v V(x,y,u,v)]
,
\\
&&
\propV_{\Fbar FF \Fbar V}(x,y,z,u,v) =
2y \bigl \lbrace
-4 U(x,y,u,v)
+ [2 B(x,y)+1] \deltaMSbar
\extranewlineforpreprintmode
+ {\mathcal L}_v [v U(x,y,u,v)]\bigr \rbrace/(y-z) +
(y \leftrightarrow z)
,
\\
&&
\propV_{\Fbar FF \Fbar V}(x,y,y,u,v) =
8y V(x,y,u,v) -8 U(x,y,u,v)
+ [4y B(x,y') + 4 B(x,y)+2] \deltaMSbar
\nonumber \\ &&
\qquad
- {\mathcal L}_v \lbrace
2v[ y  V(x,y,u,v) - U(x,y,u,v)] \rbrace
,
\\
&& \propV_{FF \Fbar \Fbar V}(x,y,z,u,v) =
\bigl \lbrace
4 (s-x-y) U(x,y,u,v) + 4 I(y,u,v)
\extranewlineforpreprintmode
+ 2 [(x+y-s) B(x,y) - A(y)+y] \deltaMSbar
\nonumber \\ && \qquad
+ {\mathcal L}_v \left [ v(x+y-s) U(x,y,u,v) - v I(y,u,v) \right ]
\bigr \rbrace/(y-z)
+ (y \leftrightarrow z)
,
\\
&& \propV_{FF \Fbar \Fbar V}(x,y,y,u,v) =
4 [(x+y-s) V(x,y,u,v) -U(x,y,u,v) + I(y',u,v)]
\extranewlineforpreprintmode
+ 2 [(x+y-s) B(x,y')
\nonumber \\ && \qquad
+ B(x,y)
- A(y)/y] \deltaMSbar
\extranewlineforpreprintmode
+ {\mathcal L}_v \left
\lbrace v[(s-x-y) V(x,y,u,v)+U(x,y,u,v) - I(y',u,v)] \right \rbrace
,
\\
&&
\propV_{\Fbar F \Fbar F V}(x,y,z,u,v) =
\Bigl [
2 (v-y-u) U(x,y,u,v) + 2 [A(v)-A(u)] B(x,y)
\extranewlineforpreprintmode
+ y[2 B(x,y) +1] \deltaMSbar
\nonumber \\ &&
\qquad
+ {\mathcal L}_v \bigl \lbrace
[ v (y+u) - (y-u)^2 ] U(x,y,u,v) + y S(x,u,v)
\extranewlineforpreprintmode
+ (u-y) A(v) B(x,y) \bigr \rbrace
\Bigr ]/(y-z)
+ (y \leftrightarrow z)
,
\phantom{xxx}
\\
&&
\propV_{\Fbar F \Fbar F V}(x,y,y,u,v) =
2 (y+u-v) V(x,y,u,v) -2 U(x,y,u,v) + 2 [A(v)-A(u)] B(x,y')
\nonumber \\ &&
\qquad
+ [2y B(x,y') + 2 B(x,y)+1] \deltaMSbar
\extranewlineforpreprintmode
+ {\mathcal L}_v \Bigl \lbrace
[ (y-u)^2 - v (y+u) ] V(x,y,u,v)
+ (2u-2y+v) U(x,y,u,v)
\nonumber \\ && \qquad
+ S(x,u,v)
+ (u-y) A(v) B(x,y')
- A(v) B(x,y) \Bigr \rbrace
,
\\
&&
\propV_{F \Fbar \Fbar F V}(x,y,z,u,v) =
\Bigl (
(y+u-v) [(s-x-y) U(x,y,u,v)
+ I(y,u,v)]
\nonumber \\ && \qquad
+ [A(u)-A(v) - y \deltaMSbar ][
(s-x-y) B(x,y)+ A(y)] \Bigr )/y(y-z) + (y \leftrightarrow z)
\extranewlineforpreprintmode
+ \Bigl ( v (s+u-x-v) T(v,x,u)
\nonumber \\ &&
\qquad
+ u (x+u-v-s) T(u,x,v)
+ (u-v) [ 2 x T(x,u,v) + 2 S(x,u,v)
\extranewlineforpreprintmode
-A(x) - A(u) - A(v) +x+u+v -s/4 ]
\Bigr )/yz
\nonumber \\ &&
\qquad
+ \deltaMSbar
+ {\mathcal L}_v \Bigl \lbrace \Bigl (
[ v (y+u) - (y-u)^2 ] [(x+y-s) U(x,y,u,v) - I(u,v,y)]
\extranewlineforpreprintmode
+ (y-u) A(v) [(s-x-y) B(x,y)
\nonumber \\ &&
\qquad
+A(y)] \Bigr )/2y(y-z) + (y \leftrightarrow z)
+ \Bigl (
u v (s+u-v-x) T(v,x,u)
\extranewlineforpreprintmode
+ u^2 (x+u-s-v) T(u,x,v)
\nonumber \\ &&
\qquad
+ u (u-v) [2 x T(x,u,v) - A(v)]
+ (y z + 2 u^2 - 2 u v) S(x,u,v)
\extranewlineforpreprintmode
+ u v [A(x)+A(u) -x-v +s/4]
\Bigr )/2yz \Bigr \rbrace
,
\\
&&
\propV_{F \Fbar \Fbar F V}(x,y,y,u,v) =
\Bigl (
y (y+u-v) [(x+y-s) V(x,y,u,v) + I(y',u,v)]
\extranewlineforpreprintmode
+ [(u-v)(x-s)-y^2] U(x,y,u,v)
\nonumber \\ && \qquad
+ (v-u) I(y,u,v)
+ [A(u)-A(v)][
y (s-x-y) B(x,y') + (x-s) B(x,y) + y]
\extranewlineforpreprintmode
+ v (s+u-x-v) T(v,x,u)
\nonumber \\ &&
\qquad
+ u (x+u-v-s) T(u,x,v)
+ (u-v) [ 2 x T(x,u,v) + 2 S(x,u,v)
\extranewlineforpreprintmode
-A(x) - A(u) - A(v) +x+u+v -s/4 ]
\Bigr )/y^2
\nonumber \\ &&
\qquad
+ [(x+y-s) B(x,y') + B(x,y) - A(y)/y] \deltaMSbar
\nonumber \\ &&
\qquad
+ {\mathcal L}_v \Bigl \lbrace
y [ v (y+u) - (y-u)^2 ] [(s-x-y) V(x,y,u,v) - I(y',u,v)]
+ [s (y^2+u v- u^2)
\nonumber \\ &&
\qquad
+ y^2 (2 u +v-x-2y) + u x (u-v) ] U(x,y,u,v)
+ (y^2 + u v - u^2) I(y,u,v)
\extranewlineforpreprintmode
+ A(v) \lbrace y(y-u)(s-x-y) B(x,y')
\nonumber \\ &&
\qquad
+ [ u (s-x) - y^2 ] B(x,y)
+ y [A(y) + y-u] \rbrace
\extranewlineforpreprintmode
+ u v (s+u-v-x) T(v,x,u)
+ u^2 (x+u-s-v) T(u,x,v)
\nonumber \\ &&
\qquad
+ u (u-v) [2 x T(x,u,v) - A(v)]
+ (y^2 + 2 u^2 - 2 u v) S(x,u,v)
\extranewlineforpreprintmode
+ u v [A(x)+A(u) -x-v +s/4]
\Bigr \rbrace/2 y^2
,
\\
&& \propV_{FFFFV}(x,y,z,u,v) =
\Bigl (
(y+u-v) [(s-x-y) U(x,y,u,v) +I(y,u,v)]
\nonumber \\ &&
\qquad
+ [ A(u)-A(v) - y \deltaMSbar ][(s-x-y) B(x,y)
+ A(y)] \Bigr )/(y-z) + (y \leftrightarrow z)
\extranewlineforpreprintmode
+ S(x,u,v) - 2 A(v)
\nonumber \\ &&
\qquad
+ [x-A(x)+y+z+2u-3s/4] \deltaMSbar
\extranewlineforpreprintmode
+ {\mathcal L}_v \Bigl \lbrace  \Bigl (
[v (y+u) - (y-u)^2] [(x+y-s) U(x,y,u,v) - I(y,u,v)]
\nonumber \\ &&
\qquad
+ (y-u) A(v) [(s-x-y) B(x,y)
+ A(y)] \Bigr )/2(y-z) + (y \leftrightarrow z) 
\extranewlineforpreprintmode
+ \Bigl (
v (3 x-3u-v+s) T(v,x,u)
\nonumber \\ &&
\qquad
+ u (3 x -u -3v+s) T(u,x,v)
+ 2 x (x-s) T(x,u,v)
\extranewlineforpreprintmode
+ (8x+3y+3z-7u-4v-2s) S(x,u,v)
\nonumber \\ &&
\qquad
+ [2 A(x) - 4 A(u) -2x+u+v] A(v)
+ v [A(x) + A(u) + x-2u-v+s/4]
\Bigr )/6 \Bigr \rbrace
,
\\
&& \propV_{FFFF V}(x,y,y,u,v) =
(y+u-v) [(x+y-s) V(x,y,u,v) +I(y',u,v)]
\extranewlineforpreprintmode
+ (v-x-2y-u+s) U(x,y,u,v)
\nonumber \\ &&
\qquad
+ I(y,u,v)
+ [ A(u)-A(v)][(s-x-y) B(x,y') - B(x,y) + \lnbar y ]
+ S(x,u,v) - 2 A(v)
\nonumber \\ &&
\qquad
+ 
[y (x+y-s) B(x,y') + (x+2y-s) B(x,y) - A(x) - 2 A(y)
+x+y+2u-3s/4] \deltaMSbar
\nonumber \\ &&
\qquad
+ {\mathcal L}_v \Bigl \lbrace
[v (y+u) - (y-u)^2] [(s-x-y) V(x,y,u,v) - I(y',u,v)]
\extranewlineforpreprintmode
+ [s (2y-2u-v) + u v + 2 u x + v x
\nonumber \\ &&
\qquad
+ 4 u y + 2 v y - 2 x y - u^2 - 3 y^2]
U(x,y,u,v)
+ [2y-2u-v] I(y,u,v)
\extranewlineforpreprintmode
+ A(v) [(y-u)(s-x-y) B(x,y')
\nonumber \\ &&
\qquad
+ (s-x-2y+u) B(x,y) -y +(2y-u) \lnbar y]
+ \Bigl (
v (3 x-3u-v+s) T(v,x,u)
\extranewlineforpreprintmode
+ u (3 x -u -3v+s) T(u,x,v)
\nonumber \\ &&
\qquad
+ 2 x (x-s) T(x,u,v)
+ (8x+6y-7u-4v-2s) S(x,u,v)
\extranewlineforpreprintmode
+ [2 A(x) - 4 A(u) -2x+u+v] A(v)
\nonumber \\ &&
\qquad
+ v [A(x) + A(u) + x-2u-v+s/4]
\Bigr )/3 \Bigr \rbrace/2 .
\end{eqnarray}
In the last case, the limit $y=z=0$ needed for massless fermions is 
slightly 
non-trivial:
\begin{eqnarray}
&& \propV_{FFFF V}(x,0,0,u,v) =
(u-v) (x-s) \Vbar (x,0,u,v) 
+ (s-x-u+v) U(x,0,u,v)
\extranewlineforpreprintmode
+ S(x,u,v) 
+ I(0,u,v)
\nonumber \\ &&
\quad
+ 2 [ A(u)-A(v)][s B(0,x')+ A(x)/x]
- 2 A(v)
\extranewlineforpreprintmode
+ 
[(x-s) B(0,x) - A(x)  +x+2u
-3s/4] \deltaMSbar
\nonumber \\ &&
\quad
+ {\mathcal L}_v \Bigl \lbrace
u (u-v) (x-s) \Vbar (x,0,u,v) 
+ [ u v + 2 u x + v x
- u^2 
- 2 u s - v s]
U(x,0,u,v)
\extranewlineforpreprintmode
-(2u+v) I(0,u,v)
\nonumber \\ &&
\quad 
+ A(v) [ (s+2u-x) B(0,x) - 2 u x B(0,x') ] 
+ \Bigl (
v (3 x-3u-v+s) T(v,x,u)
\extranewlineforpreprintmode
+ u (3 x -u -3v+s) T(u,x,v)
\nonumber \\ &&
\quad
+ 2 x (x-s) T(x,u,v)
+ (8x
-7u
-4v-2s) S(x,u,v)
\extranewlineforpreprintmode
+ [2 A(x) - 4 A(u) -2x-5u+v] A(v)
\nonumber \\ &&
\quad
+ v [A(x) + A(u) + x-2u-v+s/4]
\Bigr )/3 \Bigr \rbrace/2 .
\phantom{xxx}
\label{VFFFFVxoouv}
\end{eqnarray}
This function is well-defined as $s \rightarrow x$, even though
$B(0,x')$ and $(x-s) \Vbar (x,0,u,v)$ are each singular in that limit.

\protect\subsection{\label{subsec:SFV}\protect{Diagrams with scalar, 
fermion, and 
vector propagators}}

In this subsection I present the results for two-loop diagrams
that contain one vector propagator and both fermion and scalar
propagators.

The contributions of diagrams of topology $\propM$ with line 1 taken to be 
a vector 
propagator, lines 2,4,5 to be fermion, and line 3 to be scalar, are:
\begin{eqnarray}
\Pi^{(2)}_{ij} &=&
2 \Bigl (
\text{Im} \bigl [ g^{ail} y^{KMj} g^{aK'}_N y^{M'Nl} M_{KK'} M_{MM'} \bigr 
]
\propM_{V\Fbar S\Fbar F} (m_a^2, m_K^2, m_l^2, m_M^2, m_N^2)
\nonumber \\ 
&&
+\text{Im} \bigl [ g^{ail} y^{KMj} g^{aK'}_{N} y_{MN'l} M_{KK'} M^{NN'} 
\bigr ]
\propM_{V\Fbar S  F \Fbar} (m_a^2, m_K^2, m_l^2, m_M^2, m_N^2)
\nonumber \\ 
&&
+\text{Im} \bigl [ g^{ail} y^{KMj} g^{aN}_{K} y^{M'N'l} M_{MM'} M_{NN'} 
\bigr ]
\propM_{VFS\Fbar \Fbar} (m_a^2, m_K^2, m_l^2, m_M^2, m_N^2)
\nonumber \\ 
&&
+\text{Im} \bigl [ g^{ail} y^{KMj} g^{aN}_K y_{MNl} \bigr ]
\propM_{VFSFF} (m_a^2, m_K^2, m_l^2, m_M^2, m_N^2)
\Bigr ) + (i \leftrightarrow j)
,
\end{eqnarray}
where
\begin{eqnarray}
&& \propM_{V\Fbar S \Fbar F}(x,y,z,u,v) =
(y-x+2z-2u+v) M(x, y, z, u, v) 
\extranewlineforpreprintmode
- B(x,z) B(y,u)
-U(x,z,u,v)
\nonumber \\ && \qquad
+ 2 U(z,x,y,v)
+ U(y,u,z,v) - 2 U(u,y,x,v)
\extranewlineforpreprintmode
+ {\mathcal L}_x \Bigl [
(x-y+v)[(s-z) M(x,y,z,u,v) + U(u,y,x,v)]
\nonumber \\ && \qquad
+ (s-z) [
U(x,z,u,v) - B(x,z) B(y,u)] + S(x,u,v) - A(x) B(y,u)
\Bigr ]
,
\\
&& \propM_{V\Fbar S F \Fbar }(x,y,z,u,v) =
(y-z-2u+v-s) M(x,y,z,u,v) - U(x,z,u,v) 
\extranewlineforpreprintmode
+ 2 U(z,x,y,v) + U(u,y,x,v)
\nonumber \\ && \qquad
- B(x,z) B(y,u)
+ {\mathcal L}_x \Bigl [
(z-s)[(y+z-v-s) M(x,y,z,u,v) - U(x,z,u,v) 
\extranewlineforpreprintmode
+ B(x,z) B(y,u)]
\nonumber \\ && \qquad
+ \lbrace
[s (x+3y-v)+
u v - u x - u y + v y - x y - y^2  - 2 y z] U(u, y, x, v)
\extranewlineforpreprintmode
+ x(s -x - u + v) T(x,u,v)
\nonumber \\ && \qquad
+ v (u + v - x-s) T(v,x,u)+ 2 u (v - x) T(u, v, x)
\extranewlineforpreprintmode
+ (y - 2 x + 2v) S(x,u,v) + (x+y-v) I(x,y,v)
\nonumber \\ && \qquad
+ (s - u - y) A(x) B(y,u)
+ A(x) [A(y) + x-v] 
\extranewlineforpreprintmode
+ x [A(u) + A(v) -x-u+s/4]
\rbrace/2 y \Bigr ]
\label{eq:MVfSFf}
,
\\
&& \propM_{VFS \Fbar \Fbar }(x,y,z,u,v) =
(x-y+2u-v-2s) M(x,y,z,u,v)
+ U(x,z,u,v)
\extranewlineforpreprintmode
 - 2 U(z,x,y,v) - U(y,u,z,v)
\nonumber \\ && \qquad
+B(x,z) B(y,u)
+ {\mathcal L}_x \Bigl [
(x+y-v) [ (s-z) M(x,y,z,u,v) + U(u,y,x,v) ]
\extranewlineforpreprintmode
+ (z-s) [U(x,z,u,v)
\nonumber \\ && \qquad
- B(x,z) B(y,u)] -S(x,u,v) + A(x) B(y,u)
\Bigr ]
,
\\
&& \propM_{VFSFF}(x,y,z,u,v) =
[s (2z-2u-v) + (2 u  - v + 2 x - y - 2 z)u - y z] M(x,y,z,u,v)
\extranewlineforpreprintmode
+ (u+z) U(x,z,u,v)
\nonumber \\ && \qquad
+ (y+2u-2s) U(u,y,x,v)
- 2 u U(y,u,z,v)  + (u+s) B(x,z) B(y,u) 
\extranewlineforpreprintmode
- S(x,u,v)
+ \lbrace
[s (x+y-v)
\nonumber \\ && \qquad
-2 u x - x^2  + v z + x z - y z] U(z,x,y,v)
+ 2 (v-y) z T(z,y,v)
\extranewlineforpreprintmode
+ y (s+v-y-z) T(y,z,v)
\nonumber \\ && \qquad
+ v (v+z-s-y) T(v,y,z)
+ (x - 2 y + 2v) S(y,z,v)
\extranewlineforpreprintmode
+ [A(v)-A(y)][ (z-s) B(x,z) -A(x)]
\nonumber \\ && \qquad
+ (y-x-v) I(x,y,v)
+(y-v) [A(y) + A(z) + A(v)-y-z-v +s/4] \rbrace/x
\nonumber \\ && \qquad
+ {\mathcal L}_x \Bigl [
(z-s) [(yz-yu+vu-vs) M(x,y,z,u,v)
\extranewlineforpreprintmode
+ (u-z) U(x, z, u, v)
+ (y-v) U(z, x, y, v)
\nonumber \\ && \qquad
+ (s-u) B(x,z) B(y,u)]
+ [s (v + x + y) + u (y-x-v) 
\extranewlineforpreprintmode
+ y (y-x -2z -v)] U(u,y,x,v)/2
\nonumber \\ && \qquad
+ \bigl [x (x+3v-3u-s) T(x,u,v)
+ v (3x+v-3u-s) T(v,x,u)
+2 u (s-u) T(u,x,v)
\nonumber \\ && \qquad
+ (4 x -3 y + 6 z -8 u + 4 v - 4 s) S(x,u,v) \bigr ]/6
+(y+u-s) A(x) B(y,u)/2
\nonumber \\ && \qquad
+ (z-s) [A(y)-A(v)] B(x,z)
+ A(x) [2 A(v)/3 - A(u)/3 - A(y)/2] 
\extranewlineforpreprintmode
+ (x-y+v) I(x,y,v)/2
\nonumber \\ && \qquad
+ [(2u-x-v) A(x)
+x \lbrace 2v +x -u-s/4 - A(u)-A(v)\rbrace ]/6 \Bigr ]
.
\end{eqnarray}

The contributions from diagrams of topology $\propV$ with
line 1 taken to be vector, lines 2,3 to be scalar, and lines 4,5
to be fermion are:
\begin{eqnarray}
\Pi^{(2)}_{ij} &=&
g^{aik} g^{ajl} \Bigl (
\re \bigl [ y^{MNk} y^{M'N'l} M_{MM'} M_{NN'} \bigr ]
\propV_{VSS\Fbar\Fbar} (m_a^2, m_k^2, m_l^2, m_M^2, m_N^2)
\nonumber \\ &&
+
\re \bigl [ y^{MNk} y_{MNl} \bigr ]
\propV_{VSSFF} (m_a^2, m_k^2, m_l^2, m_M^2, m_N^2)
\Bigr )
,
\end{eqnarray}
where
\begin{eqnarray}
&& \propV_{VSS\Fbar \Fbar }(x,y,z,u,v) = -2 \propV_{VSSSS}(x,y,z,u,v)
,
\\
&& \propV_{VSSFF}(x,y,z,u,v) = \Bigl \lbrace
(y-u-v)
[(x-2y-2s) U(x,y,u,v) - I(y,u,v)]
+2y S(x,u,v)
\nonumber \\ && \quad
+ [A(u)+A(v)] [(x-2y-2s) B(x,y) -A(y)]
\Bigr \rbrace/(y-z) + (y \leftrightarrow z)
\nonumber \\ && \qquad
+ {\mathcal L}_x
\Bigr \lbrace
(y-s)^2 \lbrace
(y-u-v) U(x,y,u,v)
+ [ A(u)+A(v) ] B(x,y) \rbrace/(y-z)
+ (y \leftrightarrow z)
\nonumber \\ && \quad
+ \lbrace
u (u -3 x + 3 v -s) T(u, x, v)
+ v (3 u -3 x + v -s) T(v, x, u)
\extranewlineforpreprintmode
+ 2 x (s-x)T(x,u,v)
+ (5 s - 5 x - 3 y 
\nonumber \\ && \quad
-3z+ 4 u + 4 v) S(x,u,v)
+ 2 [A(u) +A(v)][x- A(x)]
\extranewlineforpreprintmode
+ (2x-u-v) A(x)
+x (s/2 -2x-u-v)
\rbrace/3
\Bigr \rbrace
,
\\
&& \propV_{VSSFF}(x,y,y,u,v) =
(y-u-v)
[(2y-x+2s) V(x,y,u,v) - I(y',u,v)]
\extranewlineforpreprintmode
+ [x - 4 y + 2u+2v-2s] U(x,y,u,v)
\nonumber \\ && \qquad
- I(y,u,v)
+2S(x,u,v)
+ [A(u)+A(v)] [(x-2y-2s) B(x,y') - 2 B(x,y) -\lnbar y]
\nonumber \\ && \qquad
+ {\mathcal L}_x
\Bigr \lbrace
(y-s)^2 \lbrace
(u+v-y) V(x,y,u,v)
+ [ A(u)+A(v) ] B(x,y') \rbrace
\extranewlineforpreprintmode
+ (y-s)(3y-2u-2v-s) U(x,y,u,v)
\nonumber \\ && \qquad
+ 2 (y-s) [A(u)+A(v) ] B(x,y)
+ \lbrace
u (u -3 x + 3 v -s) T(u, x, v)
\extranewlineforpreprintmode
+ v (3 u -3 x + v -s) T(v, x, u)
\nonumber \\ && \qquad
+ 2 x (s-x)T(x,u,v)
+ (5 s - 5 x - 3 y -3z+ 4 u + 4 v) S(x,u,v)
\extranewlineforpreprintmode
+ 2 [A(u) +A(v)][x- A(x)]
\nonumber \\ && \qquad
+ (2x-u-v) A(x)
+x (s/2 -2x-u-v)
\rbrace/3
\Bigr \rbrace
.
\end{eqnarray}

Finally, the results for diagrams of topology $\propV$ with lines 1,3 
taken to be
scalars, 2 to be vector, and 4,5 to be fermion, are:
\begin{eqnarray}
&& \Pi^{(2)}_{ij} \>=\>
2 \text{Im} \bigl [(g^{aik} \lambda^{jkl} + g^{ajk} \lambda^{ikl})
g^{aN}_{M} y^{MN'l} M_{NN'} \bigr ]
\propV_{SVSF\Fbar} (m_k^2, m_a^2, m_l^2, m_M^2, m_N^2)
\end{eqnarray}
where
\begin{eqnarray}
&& \propV_{SVSF \Fbar }(x,y,z,u,v) =
\Bigl (
(y+u-v) [(s-x) U(x,y,u,v) + I(y,u,v)]
\extranewlineforpreprintmode
+ [A(u) - A(v)][(s-x) B(x,y)
\nonumber \\ && \qquad
+ A(y)]
\Bigr )/y (y-z) + (y \leftrightarrow z)
+
\extranewlineforpreprintmode
\Bigl (
u (x+u-v-s) T(u,x,v)
+v (s+u-x-v) T(v,x,u)
\nonumber \\ && \qquad
+ (u-v) [2 x T(x,u,v) +2 S(x,u,v) - A(x)
-A(u)-A(v)
+x+u+v-s/4]
\Bigr )/yz
\nonumber \\ && \qquad
+
{\mathcal L}_y
\Bigl \lbrace \Bigl (
(v-y-u)[(s-x) U(x,y,u,v) + I(y,u,v)]
\extranewlineforpreprintmode
+ [A(v)-A(u)][(s-x) B(x,y) + A(y)] \Bigr )/(y-z)
\nonumber \\ && \qquad
+ (y \leftrightarrow z)
\Bigr \rbrace
,
\\
&& \propV_{SVSF \Fbar }^{\xi \not= 1}(x,y,y,u,v) =
\Bigl (
(v-y-u)[(s-x) U(x,y,u,v) + I(y,u,v)]
\extranewlineforpreprintmode
+ [A(v)-A(u)][(s-x) B(x,y) + A(y)]
\nonumber \\ && \qquad
+ u (x+u-v-s) T(u,x,v)
+v (s+u-x-v) T(v,x,u)
\extranewlineforpreprintmode
+ (u-v) [2 x T(x,u,v)
+2 S(x,u,v) 
\nonumber \\ && \qquad
- A(x)
-A(u)-A(v)
+x+u+v-s/4]
\Bigr )/y^2
\nonumber \\ && \qquad
+ \frac{1}{(1-\xi) y} 
{\mathcal L}_y
 \Bigl \lbrace
(y+u-v) [(s-x) U(x,y,u,v) + I(y,u,v)]
\extranewlineforpreprintmode
+ [A(u) - A(v)][(s-x) B(x,y)
+ A(y)]
\Bigr \rbrace 
,
\phantom{xxx}
\\
&& \propV_{SVSF \Fbar }^{\xi=1}(x,y,y,u,v) =
\Bigl \lbrace
y (y+u-v) [(x-s) V(x,y,u,v) + I(y',u,v)]
\extranewlineforpreprintmode
+ u (x+u-v-s) T(u,x,v)
\nonumber \\ && \qquad
+v (s+u-x-v) T(v,x,u)
+ (u-v) [
(x-s) U(x,y,u,v) - I(y,u,v)
\extranewlineforpreprintmode
+ 2 x T(x,u,v)
+2 S(x,u,v)
- A(x)
\nonumber \\ && \qquad
-A(u)-A(v)
+x+u+v-s/4]
\extranewlineforpreprintmode
+ [A(u) - A(v)][(s-x) y B(x,y')
+ (x-s) B(x,y) +y ]
\Bigr \rbrace/y^2
.
\end{eqnarray}

\protect\section{\label{sec:massless}\protect{Two-loop functions involving 
massless vectors}}
\setcounter{equation}{0}

In the preceding formulas, the vector bosons have been taken to have
generic non-zero masses. However, when the corresponding gauge symmetry is
unbroken, the vector boson will be massless. Also, the three-point
couplings of such massless vector bosons occur only when the other two
particles are degenerate in mass.  In this section I will give the
appropriate limits of the two-loop integral functions for massless
vector bosons. It should be noted that the presentations of some of the
results in this section are not unique, because of the existence of
identities involving basis functions with vanishing squared masses [see 
equations (\ref{eq:defVbarxozu})-(\ref{eq:SOOx}) in the 
Appendix].

For the integral functions of subsection \ref{subsec:SV}, one finds:
\begin{eqnarray}
\propW_{SSSV}(x,x,x,0) &=& 3 I(0,x,x) - A(x) + 2x
,
\\
\propX_{SSV}(x,x,0) &=& 0
,
\\
\propY_{SSSV}(x,y,y,0) &=& 0
,
\\
\propY_{VSSS}(0,y,z,u) &=&
A(u) \bigl \lbrace
(3-\xi)(y+s) B(0,y) + (3 - 2 \xi) A(y) \bigr \rbrace/(y-z)
\extranewlineforpreprintmode
+ (y \leftrightarrow z)
,
\\
\propY_{VSSS}(0,y,y,u) &=&
A(u) \bigl \lbrace
(3 - \xi)[(y+s) B(0,y') + B(0,y)]
+ (3 - 2 \xi) \lnbar y
\bigr \rbrace
,
\\
\propU_{SVSS}(x,0,z,z) &=& 0
,
\\
\propV_{SSSSV}(x,y,y,y,0) &=&
4y V(x,y,y,0)
-2 U(x,y,y,0)
-A(y) B(x,y')
+(1-\xi) \Tbar(0,x,y)
,
\\
\propV_{VSSSS}(0,y,z,u,v) &=& [(\xi-3) (y+s) U(0,y,u,v)
+ (2 \xi-3) I(y,u,v)]/(y-z) 
\extranewlineforpreprintmode
+ (y \leftrightarrow z)
,
\\
\propV_{VSSSS}(0,y,y,u,v) &=& (3-\xi) [(y+s) V(0,y,u,v)
- U(0,y,u,v) ] + (2 \xi-3) I(y',u,v)]
,
\\
\propV_{SVSSS}(x,0,z,u,u) &=& 0
,
\\
\propM_{SSSSV}(x,x,y,y,0) &=&
2 (x+y-s) M(x,x,y,y,0) -2 U(x,y,y,0) -2 U(y,x,x,0)
+B(x,y)^2
\nonumber \\ &&
+(2\xi-2) \Tbar(0,x,y)
,
\\
\propM_{VSSSS}(0,y,z,u,y) &=&
(2y+z-2u+s) M(0,y,z,u,y) + U(y,u,y,z)
\extranewlineforpreprintmode
-U(u,y,0,y) - U(0,z,y,u)
\nonumber \\ &&
+ 2 U(z,0,y,y) - B(0,z) B(y,u)
\extranewlineforpreprintmode
+ (1-\xi) \bigl \lbrace
(z+s) [U(0,z,y,u) - B(0,z) B(y,u)]
\nonumber \\ &&
- 2 y T(y,0,u) - 2 u T(u,0,y)
- 4 S(0,y,u)
\extranewlineforpreprintmode
+ 2 I(y,z,u) + 2 [s-A(z)] B(y,u)
\nonumber \\ &&
+ 2 A(u) + 2 A(y)-2u-2y+s/2
\bigr \rbrace/(z-s)
.
\end{eqnarray}

The massless-vector limits of the integrals of subsection 
\ref{subsec:FV} are:
\begin{eqnarray}
\propM_{\Fbar \Fbar \Fbar \Fbar V}(x,x,y,y,0) &=&
(6 + 2\xi) M (x, x, y, y, 0)
,
\\
\propM_{F\Fbar\Fbar FV}(x,x,y,y,0) &=&
(1+\xi)[(x+y-s) M(x,x,y,y,0) - U(x,y,y,0) - U(y,x,x,0)]
\nonumber \\ &&
+2 (1-\xi) \Tbar (0,x,y)
+ 2 B(x,y)^2 + 2 \deltaMSbar
,
\\
\propM_{F\Fbar F\Fbar V}(x,x,y,y,0) &=&
(3+\xi)[(x+y-s) M(x,x,y,y,0) - U(x,y,y,0) - U(y,x,x,0)]
\nonumber \\ &&
+ [4 B(x,y)+2] \deltaMSbar
,
\\
\propM_{FF \Fbar \Fbar V}(x,x,y,y,0) &=&
2 x (1 + \xi) M(x,x,y,y,0)
+ [(1-\xi)(x+y-s)/y -4] U(x,y,y,0)
\extranewlineforpreprintmode
+ 2 B(x,y)^2
\nonumber \\ &&
+ (1-\xi)
[4 x T(x,0,y) +2 y T(y,0,x) 
\extranewlineforpreprintmode
+(x+y-s) \Tbar(0,x,y)
+4 S(0,x,y)
\nonumber \\ &&
-I(0,y,y)
-2 A(x) -4 A(y) + 3 (x+y-s/2)]/y
+ 2 \deltaMSbar
,
\\
\propM_{FFFFV}(x,x,y,y,0) &=&
2 [(x+y-s)^2 +(\xi-1) x y] M(x,x,y,y,0)
\extranewlineforpreprintmode
+ (3 + \xi) (s-x-y) [U(x,y,y,0)
\nonumber \\ &&
+ U(y,x,x,0)]
+ 2 (1-\xi) [ x T(x,0,y) + y T(y,0,x) - A(x)
\extranewlineforpreprintmode
-A(y) +x+y-s/2]
\nonumber \\ &&
+ (6-2\xi) S(0,x,y)
+ (1+\xi)[I(0,x,x)+I(0,y,y)]
\nonumber \\ &&
+ 4 [(x+y-s) B(x,y) -A(x) - A(y) 
\extranewlineforpreprintmode
+x+y-s/4] \deltaMSbar
,
\end{eqnarray}
and
\begin{eqnarray}
&&
\propV_{\Fbar\Fbar\Fbar\Fbar V}(x,y,y,y,0) \>=\>
(6+2\xi) V(x,y,y,0)
+ 4 B(x,y') \deltaMSbar
,
\\
&&
\propV_{\Fbar FF \Fbar V}(x,y,y,y,0) \>=\>
(6+2 \xi)[yV(x,y,y,0) - U(x,y,y,0)]
\extranewlineforpreprintmode
+ [4 y B(x,y') + 4 B(x,y)+2] \deltaMSbar
,
\\
&&
\propV_{FF \Fbar \Fbar V}(x,y,y,y,0) \>=\>
(3+\xi) [(x+y-s) V(x,y,y,0) - U(x,y,y,0) -A(y)^2/2y^2]
\nonumber \\ &&
\qquad\qquad
+ 2 [(x+y-s) B(x,y') + B(x,y) - A(y)/y] \deltaMSbar
,
\\
&&
\propV_{\Fbar F \Fbar F V}(x,y,y,y,0) \>=\>
2 \xi \Tbar(0,x,y) + 2 [\xi A(y) - (1+\xi) y] B(x,y')
-(1+\xi) [ 1+ 2 B(x,y) ]
\nonumber \\ &&
\qquad\qquad
+ [2 y B(x,y') + 2 B(x,y)+1] \deltaMSbar
,
\\
&&
\propV_{F\Fbar\Fbar F V}(x,y,y,y,0) \>=\>
\Bigl [
\xi \bigl \lbrace (x+y-s) \Tbar(0,x,y) + (s-x+y) T(y,0,x)
\extranewlineforpreprintmode
+2 x T(x,0,y) + 2 S(0,x,y) - A(x)
\nonumber \\ &&
\qquad\qquad
+ 2x+2y -5s/4 \bigr \rbrace
+(x+y-s)[\xi A(y) - (1+\xi) y] B(x,y')
\extranewlineforpreprintmode
+[\xi (s-x) A(y)/y - (1+\xi) y] B(x,y)
\nonumber \\ &&
\qquad\qquad
+(1- 3 \xi) A(y)
\Bigr ]/y
+ [(x+y-s) B(x,y') + B(x,y) -A(y)/y] \deltaMSbar
,
\\
&&
\propV_{FFFFV}(x,y,y,y,0) \>=\>
\xi \bigl [(x+y-s) \Tbar(0,x,y)
+ 2 y T(y,0,x) + S(0,x,y)
+ I(0,y,y)
\bigr ]
\nonumber \\ &&
\qquad\qquad
+(x+y-s) [\xi A(y) - (1+\xi) y ] B(x,y')
\extranewlineforpreprintmode
+[\xi A(y) + (1+\xi)(s-x-2y)] B(x,y)
+ (1- \xi) A(x)
\nonumber \\ &&
\qquad\qquad
+ (2 - \xi) A(y)
-x-3 y +(3 + \xi) s/4
\extranewlineforpreprintmode
+ [ y (x+y-s) B(x,y') + (x+2y-s) B(x,y) -A(x) 
\nonumber \\ &&
\qquad\qquad
- 2 A(y) 
+x+3y-3s/4] 
\deltaMSbar .
\end{eqnarray}

Finally, the integrals introduced in subsection \ref{subsec:SFV} have
the following forms for massless vectors:
\begin{eqnarray}
&& \propM_{V\Fbar S \Fbar F} (0,y,z,u,y) =
2 (y+z-u) M(0,y,z,u,y)
+ U(y,u,y,z)
\extranewlineforpreprintmode
- U(0,z,y,u)
+ 2 U(z,0,y,y)
\nonumber \\ && \qquad\qquad
- 2 U(u,y,0,y)
- B(0,z) B(y,u)
+ (1-\xi) \Bigl [ (s-z) M(0,y,z,u,y) + U(u,y,0,y)
\nonumber \\ && \qquad\qquad
+ \bigl \lbrace (z+s) U(0,z,y,u)
- 2 y T(y,0,u)
- 2 u T(u,0,y) - 4 S(0,y,u)
+ 2 I(y,z,u)
\nonumber \\ && \qquad\qquad
-(z+s) B(0,z) B(y,u)
+ 2 [s - A(z)] B(y,u)
\extranewlineforpreprintmode
+ 2 A(y) + 2 A(u)
-2u-2y+s/2 \bigr \rbrace/(z-s) \Bigr ]
,
\\
&& \propM_{VFS \Fbar \Fbar} (0,y,z,u,y) =
-\propM_{V\Fbar S \Fbar F} (0,y,z,u,y) + 2 \xi [
(z-s) M(0,y,z,u,y) 
\extranewlineforpreprintmode
- U(u,y,0,y) ]
,
\\
&& \propM_{V\Fbar S F\Fbar} (0,y,z,u,y) =
(2y-z-2u-s) M(0,y,z,u,y)
- U(0,z,y,u)
\extranewlineforpreprintmode
+2 U(z,0,y,y)
+ U(u,y,0,y)
\nonumber \\ && \qquad\qquad
- B(0,z) B(y,u)
+ (1-\xi) \Bigl [
S(y,y,z) - u U(y,u,y,z)
\extranewlineforpreprintmode
+[A(u) - A(y) -z/2 - s/2] B(0,z)
\nonumber \\ && \qquad\qquad
-A(z)/2
+ \Bigl \lbrace
(u z + y s-z s) U(0,z,y,u)
\extranewlineforpreprintmode
+ (s-y-u) [ y T(y,0,u) + u T(u,0,y)
+ 2 S(0,y,u)
\nonumber \\ && \qquad\qquad
+ \lbrace A(z)-s \rbrace B(y,u) - A(u) ]
+ (y-z+u) I(y,z,u)
\extranewlineforpreprintmode
+ (s z - s u - y z) B(0,z) B(y,u)
\nonumber \\ && \qquad\qquad
+ (y+u-s) A(y) - 5 s^2/8 + (10 u + 10 y + 3 z)s/8
- (y+u)^2 \Bigr \rbrace/(z-s)
\Bigr ]/y
,
\label{eq:MVfSFfzero}
\\
&& \propM_{VFSFF} (0,y,z,u,y) =
[s (2z-y-2u) -y z + 2 u (u-y-z)] M(0,y,z,u,y)
\extranewlineforpreprintmode
- 2 u U(y,u,y,z)
\nonumber \\ && \qquad\qquad
+ (u+z) U(0,z,y,u)
+ (s-2u+z) U(z,0,y,y)
\extranewlineforpreprintmode
+ (y+2u-2s) U(u,y,0,y) + S(y,y,z) - S(0,y,u)
\nonumber \\ && \qquad\qquad
-I(0,y,y)
+(u+s) B(0,z) B(y,u)
+ (1-\xi) \Bigl [
S(y,y,z) -u U(y,u,y,z)
-A(z)/2
\nonumber \\ && \qquad\qquad
+ [A(u)-A(y)-z/2-s/2] B(0,z)
+ \Bigl\lbrace
(z^2 - y z - s u) U(0,z,y,u) 
\extranewlineforpreprintmode
+(y+u-s) [
y T(y,0,u)
\nonumber \\ && \qquad\qquad
+ u T(u,0,y) + \lbrace A(z)-s \rbrace B(y,u) - A(u) ]
\extranewlineforpreprintmode
+(2 u +2 y -z -s) S(0,y,u)
+ (z-y-u) I(y,z,u)
\nonumber \\ && \qquad\qquad
+ (z u + y s - s^2) B(0,z) B(y,u)
+ (s-y-u) A(y)
+ 3 s^2/8 -s (10 y +10 u +z)/8
\nonumber \\ && \qquad\qquad
+ (y+u)^2
\Bigr \rbrace/(z-s) \Bigr ]
,
\end{eqnarray}
and
\begin{eqnarray}
&& \propV_{VSSFF} (0,y,z,u,v) =
\Bigl [
(\xi-3) (y+s) [(y-u-v) U(0,y,u,v) + \lbrace A(u) + A(v) \rbrace B(0,y)]
\nonumber \\ && \qquad\qquad
+ (2 \xi - 3) [(y-u-v) I(y,u,v) + \lbrace A(u) + A(v) \rbrace A(y)]
\Bigr ]/(y-z) + (y \leftrightarrow z)
\nonumber \\ && \qquad\qquad
+ (3 - \xi) S(0,u,v) + (1-\xi) s/2
,
\\
&& \propV_{VSSFF} (0,y,y,u,v) =
(3-\xi) \bigl \lbrace (y+s)(y-u-v) V(0,y,u,v)
\extranewlineforpreprintmode
+(u+v-2y-s) U(0,y,u,v) + S(0,u,v)
\nonumber \\ && \qquad\qquad
- [ A(u) + A(v) ] [(y+s) B(0,y') + B(0,y)] \bigr \rbrace
\extranewlineforpreprintmode
+ (2 \xi - 3) \bigl [(y-u-v) I(y',u,v) +I(y,u,v)
\nonumber \\ && \qquad\qquad
+ \lbrace A(u) + A(v) \rbrace \lnbar y \bigr ]
+ (1-\xi) s/2
,
\\
&&
\propV_{SVSF\Fbar} (x,0,z,u,u) =
\xi \Bigl [ (s-x) [U(x,z,u,u) - U(x,0,u,u)] 
\extranewlineforpreprintmode
+ I(z,u,u) - I(0,u,u)
\Bigr ]/z
.
\end{eqnarray}

Certain combinations of the above functions involving massless vector 
bosons often appear together,
and will be noted for future reference. In the self-energy functions of 
scalars that are neutral
under the gauge symmetry corresponding to the massless vector boson,
the combinations
\begin{eqnarray}
\propgaugeSS (x,y) &=& \propM_{SSSSV}(x,x,y,y,0)
+\propV_{SSSSV}(x,y,y,y,0)
+\propV_{SSSSV}(y,x,x,x,0)
\nonumber \\
&=& 2 (x+y-s) M(x,x,y,y,0)
+4y V(x,y,y,0)
+4x V(y,x,x,0)
-4 U(x,y,y,0)
\nonumber \\ &&
-4 U(y,x,x,0)
-A(y) B(x,y') -A(x) B(x',y) +B(x,y)^2
,
\label{eq:defpropgaugeSS}
\\
\propgaugeFF (x,y) &=& \Bigl [
\propV_{FFFFV} (x,y,y,y,0)
+ y \propV_{F\Fbar\Fbar FV} (x,y,y,y,0)
-2 y \propV_{FF\Fbar\Fbar V} (x,y,y,y,0)
\nonumber \\ &&
- \propM_{FFFFV} (x,x,y,y,0)/2
+ y \propM_{FF\Fbar\Fbar V}(x,x,y,y,0)
\extranewlineforpreprintmode
- x y \propM_{\Fbar\Fbar\Fbar\Fbar V}(x,x,y,y,0)/2 \Bigr ]
\extranewlineforpreprintmode
+ (x \leftrightarrow y)
\label{eq:defpropgaugeFF}
\nonumber \\
&=& \Bigl [
- (x+y-s)^2 M(x,x,y,y,0)
+ 4 (x+y-s)[U(x,y,y,0) 
\extranewlineforpreprintmode
-y V(x,y,y,0)] +S(0,x,y)
\nonumber \\ &&
+ (3x+y-s) T(x,0,y) 
+ (x+y) B(x,y)^2 
%%% Typo alert! Previous line was incorrectly given in v1 and v2 as: 
%%% + (x+y)^2  B(x,y)^2 
+ (y-x-s) A(x) B(x,y)/x
\nonumber \\ &&
+ 2 (s-x-y) A(x) B(y,x') - 5 I(0,x,x) + 6 A(x) - 4 x -s/4
\nonumber \\ &&
+ \deltaMSbar \bigl \lbrace
2 y (s-x-y) B(x,y') + (s-4x) B(x,y) 
+ 4 A(x)  + 2 x  - s/4 \bigr \rbrace
\Bigr ] 
+ (x \leftrightarrow y)
,
\\
\propgaugeFbarFbar (x,y) &=& \Bigl [
2 \propV_{\Fbar F \Fbar FV}(x,y,y,y,0)
- \propV_{\Fbar FF \Fbar V}(x,y,y,y,0)
-y \propV_{\Fbar\Fbar\Fbar\Fbar V}(x,y,y,y,0)
\nonumber \\ &&
+ \propM_{F\Fbar\Fbar F V}(x,x,y,y,0)
- \propM_{F\Fbar F\Fbar V}(x,x,y,y,0)
\Bigr ] + (x \leftrightarrow y)
\label{eq:defpropgaugeFbarFbar}
\nonumber \\ &=&
\Bigl [(2s-4x) M(x,x,y,y,0) + 8 U(x,y,y,0) -8 y V(x,y,y,0)
\extranewlineforpreprintmode
+ 2 B(x,y)^2 - 4 A(x) B(y,x') 
\nonumber \\ &&
+ \deltaMSbar \bigl \lbrace
-4 B(x,y) - 4 y B(x,y') \bigr \rbrace
\Bigr ]
+ (x \leftrightarrow y)
\end{eqnarray}
will appear. They are each independent of the gauge parameter $\xi$.
For scalars that transform non-trivially under the gauge symmetry
corresponding to the massless vector boson, the 
following combinations also occur:
\begin{eqnarray}
\propgaugeSSSS (x,y,z,u) &=&
\propV_{SSSSV}(z,u,u,u,0) 
+\propV_{VSSSS}(0,x,y,z,u)
+\propM_{VSSSS}(0,u,x,z,u)
\nonumber \\ &&
+\propM_{VSSSS}(0,u,y,z,u)
,
\\
\propgaugeSSFF (x,y,z,u) &=& 
\propV_{VSSFF}(0,x,y,z,u)
+ \propV_{FFFFV}(z,u,u,u,0) 
+ u \propV_{F\Fbar\Fbar FV}(z,u,u,u,0)
\nonumber \\ &&
-2 u \propV_{FF\Fbar\Fbar V}(z,u,u,u,0)
+ \propM_{VFSFF}(0,u,x,z,u)
\extranewlineforpreprintmode
+ \propM_{VFSFF}(0,u,y,z,u)
\nonumber \\ &&
- u \propM_{V\Fbar SF\Fbar}(0,u,x,z,u)
- u \propM_{V\Fbar SF\Fbar}(0,u,y,z,u)
,
\\
\propgaugeSSFbarFbar (x,y,z,u) &=& 
\propV_{VSS\Fbar\Fbar}(0,x,y,z,u)
- \propV_{\Fbar FF\Fbar V}(z,u,u,u,0) 
+ 2 \propV_{\Fbar F\Fbar FV}(z,u,u,u,0)
\nonumber \\ &&
-u \propV_{\Fbar\Fbar\Fbar\Fbar V}(z,u,u,u,0)
+ \propM_{VFS\Fbar\Fbar}(0,u,x,z,u)
\extranewlineforpreprintmode
+ \propM_{VFS\Fbar\Fbar}(0,u,y,z,u)
\nonumber \\ &&
- \propM_{V\Fbar S\Fbar F}(0,u,x,z,u)
- \propM_{V\Fbar S\Fbar F}(0,u,y,z,u)
.
\end{eqnarray}
These are not gauge-invariant, but satisfy:
\begin{eqnarray}
\frac{\partial}{\partial \xi} \propgaugeSSSS (x,x,y,z) &=&
-\propB_{SS}(y,z) 
\frac{\partial}{\partial s} \frac{\partial}{\partial \xi} \propB_{SV}(x,0) 
+ \ldots
,
\\
\frac{\partial}{\partial \xi} \propgaugeSSFF (x,x,y,z) &=&
-\propB_{FF}(y,z) 
\frac{\partial}{\partial s} \frac{\partial}{\partial \xi} \propB_{SV}(x,0) 
+ \ldots
,
\\
\frac{\partial}{\partial \xi} \propgaugeSSFbarFbar (x,x,y,z) &=&
-\propB_{\Fbar\Fbar}(y,z) 
\frac{\partial}{\partial s} \frac{\partial}{\partial \xi} \propB_{SV}(x,0) 
+ \ldots
,
\end{eqnarray}
where, from equation 
(\ref{eq:BSVzero}) and the explicit forms of $A(x)$ and $B(0,x)$
given in equations (2.10) and (6.4) of \cite{Martin:2003qz},
\begin{eqnarray}
\frac{\partial}{\partial s} 
\frac{\partial}{\partial \xi}
\propB_{SV}(x,0)
= 2 + 2 \lnbar (x-s) - \lnbar x + \ldots .
\end{eqnarray}
The ellipses in each case 
refer to terms that vanish as $s \rightarrow x$.
These identities are useful for checking gauge invariance of the pole 
mass.
\end{widetext}

\protect\section{\label{sec:examples}\protect{A supersymmetric example}}
\setcounter{equation}{0}

In this section, I will work through the details of an example
intended to serve both as a consistency check and as a 
point of reference for the preceding results.

Consider a supersymmetric model with three chiral superfields 
$\Phi_0$, $\Phi_+$, and $\Phi_-$, with charges $0,+1,-1$ under a $U(1)$ 
gauge symmetry. 
The complex scalar components of the superfields can
be written in terms of real scalars $R_{0,+,-}$ and $I_{0,+,-}$
as
$\phi_0 = (R_0 + i I_0)/\sqrt{2}$ and
$\phi_+ = (R_+ + i I_+)/\sqrt{2}$ and
$\phi_- = (R_- + i I_-)/\sqrt{2}$,
and their two-component Weyl fermionic components are
$\psi_0$, $\psi_+$, and $\psi_-$ respectively.
Also, there is a $U(1)$ gauge field $A^\mu$, and a 
gaugino two-component Weyl 
fermion $\chi$. These fields couple to the charged fields with strength 
$g$. The superpotential is given by:
\begin{eqnarray}
W = 
Y \Phi_0 \Phi_+ \Phi_- 
+ \frac{1}{2} M_0 \Phi_0^2 + {M_{\pm}} \Phi_+ \Phi_-
.
\end{eqnarray}
Let us take $M_0$, $M_\pm$, and $Y$ to be real and positive.
In the following, I will denote $x = M_0^2$ and $z = M_{\pm}^2$.
Then the real scalars $R_0$, $I_0$, $R_+$, $I_+$, $R_-$, $I_-$ have 
squared masses $x$, $x$, $z$, $z$, $z$, $z$; the Weyl fermions 
$\psi_0$, $\psi_+$, $\psi_-$, $\chi$
have masses $\sqrt{x}$, $\sqrt{z}$, $\sqrt{z}$, and $0$; and the
gauge field $A$ is massless.
The interaction Lagrangian density is given by:
\begin{widetext}
\begin{eqnarray}
-{\cal L} &=& 
Y \sqrt{x} (\phi_+ \phi_- \phi_0^* + \text{c.c.})
+ Y \sqrt{z} (\phi_0 + \phi_0^*) (|\phi_+|^2 + |\phi_-|^2)
\extranewlineforpreprintmode
+ Y^2 \left ( 
|\phi_+|^2|\phi_-|^2 +
|\phi_+|^2|\phi_0|^2 +
|\phi_-|^2|\phi_0|^2 \right )
\nonumber \\ &&
+ g^2 (|\phi_+|^2 - |\phi_-|^2)^2/2
+ Y \left [
(\phi_0 \psi_+  \psi_-
+\phi_+ \psi_-  \psi_0
+\phi_- \psi_+  \psi_0) + \text{c.c.} \right ]
\extranewlineforpreprintmode
+ \sqrt{2} g \left [ (\phi_+^* \chi \psi_+  - \phi_-^*\chi \psi_-) +
\mbox{c.c.} \right ]
\nonumber \\ &&
+i g A^\mu (
\phi_+ \partial_\mu \phi_+^*
-\phi_+^* \partial_\mu \phi_+
+ \phi_-^* \partial_\mu \phi_-
-\phi_- \partial_\mu \phi_-^*
)
+ g^2 (|\phi_+|^2 + |\phi_-|^2) A^\mu A_\mu 
\nonumber \\ &&
- g A^\mu [
\psi_+^\dagger \overline \sigma_\mu \psi_+
-\psi_-^\dagger \overline \sigma_\mu \psi_- ]
.
\end{eqnarray}
This determines the couplings of the theory as specified
in eqs.~(\ref{LS})-(\ref{LFV}), for example:
\begin{eqnarray}
&&\lambda^{R_+R_+R_-R_-} = Y^2 - g^2,
\qquad
\lambda^{R_0R_+R_-} = Y\sqrt{x/2},
\qquad
\extranewlineforpreprintmode
y^{\psi_+ \chi I_+} = -ig,
\qquad
g^{AR_+I_+} = - g^{AI_+R_+} = g,
\phantom{xxx}
\end{eqnarray}
etc.

Applying the results of section \ref{sec:oneloop} to this model,
one obtains for the one-loop self-energy functions of the neutral
scalars:
\begin{eqnarray}
\Pi_{I_0I_0}^{(1)} &=&
Y^2 [ 2 \propA_S(z) + x \propB_{SS}(z,z)
+ \propB_{FF}(z,z)
- z \propB_{\Fbar\Fbar}(z,z)] ,
\label{eq:pixIone}
\\
\Pi_{R_0R_0}^{(1)} &=&
Y^2 [ 2 \propA_S(z) + (x +4 z) \propB_{SS}(z,z)
+ \propB_{FF}(z,z)
+ z \propB_{\Fbar\Fbar}(z,z)] ,
\label{eq:pixRone}
\\
\Pi_{I_0R_0}^{(1)} &=& 0
\end{eqnarray}
It follows immediately from equations (\ref{eq:defBSS}) and 
(\ref{eq:defBFbarFbar}) that the difference between $\Pi_{I_0I_0}^{(1)}$
and $\Pi_{R_0R_0}^{(1)}$
vanishes identically. Since there is also no mixing in the
one-loop self-energy matrix between $R_0$ and $I_0$, one can write a single 
self-energy function for the neutral complex scalar field.
Putting equation (\ref{eq:pixIone}) [or (\ref{eq:pixRone})] in terms of 
the one-loop basis functions gives simply:
\begin{eqnarray}
\Pi_{\phi_0}^{(1)} = -Y^2 (x+s) B(z,z) .
\label{eq:pineutralone}
\end{eqnarray}
At two-loop order, we have, by direct application of the results of 
sections 
\ref{sec:twoloop} and \ref{sec:massless}:
\begin{eqnarray}
\Pi_{I_0I_0}^{(2)} &=& 2 Y^4 [
\propS_{SSS}(x, z, z)
+ \propX_{SSS}(z, z, x)
+ \propX_{SSS}(z, z, z)
+ (x + 2z) \propW_{SSSS}(z, z, x, z)
\nonumber \\ && \quad
+ 2 x \propU_{SSSS}(z, z, x, z)
+ x \propY_{SSSS}(z, z, z, x)
\extranewlineforpreprintmode
+ x \propY_{SSSS}(z, z, z, z)
+ x \propZ_{SSSS}(z, z, z, z)/2
\nonumber \\ && \quad
+ (x^2 + 2 x z) \propV_{SSSSS}(z, z, z, x, z)
+ x z \propM_{SSSSS}(z, z, z, z, x)
+ \propW_{SSFF}(z, z, x, z)
\nonumber \\ && \quad
+ x \propV_{SSSFF}(z, z, z, x, z)
- 2 z \propV_{\Fbar F\Fbar FS}(z, z, z, x, z)
\extranewlineforpreprintmode
- 2 z \propV_{\Fbar F\Fbar FS}(z, z, z, z, x)
+ z \propV_{F\Fbar\Fbar FS}(z, z, z, x, z)
\nonumber \\ && \quad
+ z \propV_{F\Fbar\Fbar FS}(z, z, z, z, x)
+ \propV_{FFFFS}(z, z, z, x, z)
\extranewlineforpreprintmode
+ \propV_{FFFFS}(z, z, z, z, x)
+ z \propM_{F\Fbar\Fbar FS}(z, z, z, z, x)
\nonumber \\ && \quad
- z \propM_{FF\Fbar\Fbar S}(z, z, z, z, x)
- x z \propM_{S\Fbar S \Fbar\Fbar}(z, z, z, z, x)
+ x \propM_{SFSF\Fbar}(z, z, z, z, x)
]
\nonumber \\ &&
+ g^2 Y^2 [
2 \propX_{SSS}(z, z, z)
+ 2x \propY_{SSSS}(z, z, z, z)
\extranewlineforpreprintmode
- x \propZ_{SSSS}(z, z, z, z)
+ 4 \propW_{SSFF}(z, z, 0, z)
\nonumber \\ && \quad
+ 4 x \propV_{SSSFF}(z, z, z, 0, z)
- 8 z \propV_{\Fbar F\Fbar FS}(z, z, z, 0, z)
\extranewlineforpreprintmode
+ 4 z \propV_{F\Fbar\Fbar FS}(z, z, z, 0, z)
+ 4 \propV_{FFFFS}(z, z, z, 0, z)
\nonumber \\ && \quad
+ 2 \propW_{SSSV}(z, z, z, 0)
+ x \propgaugeSS (z,z) + \propgaugeFF (z,z) - z \propgaugeFbarFbar (z,z)
]
,
\label{eq:piIItwo}
\end{eqnarray}
and
\begin{eqnarray}
\Pi_{R_0R_0}^{(2)} &-& \Pi_{I_0I_0}^{(2)}
= 4 Y^4 z [
4 \propU_{SSSS}(z, z, x, z)
+ 2 \propY_{SSSS}(z, z, z, x)
\extranewlineforpreprintmode
+ 2 \propY_{SSSS}(z, z, z, z)
+ \propZ_{SSSS}(z, z, z, z)
\nonumber \\ && \quad
+ (6 x + 4z) \propV_{SSSSS} (z, z, z, x, z)
+ (3 x +2 z) \propM_{SSSSS}(z, z, z, z, x)
\extranewlineforpreprintmode
+ 2 \propV_{SSSFF}(z, z, z, x, z)
\nonumber \\ && \quad
+ 2 x \propV_{SSS\Fbar\Fbar}(z, z, z, x, z)
+ 2 \propV_{\Fbar F \Fbar FS}(z, z, z, x, z)
+ 2 \propV_{\Fbar F \Fbar FS}(z, z, z, z, x)
\nonumber \\ && \quad
+ \propM_{FF\Fbar\Fbar S} (z, z, z, z, x)
+ x \propM_{S\Fbar S\Fbar\Fbar} (z, z, z, z, x)
+ 2 \propM_{SFS\Fbar F} (z, z, z, z, x)
]
\nonumber \\ &&
+ 4 g^2 Y^2 z [
\propZ_{SSSS} (z, z, z, z)
+ 2 \propY_{SSSS} (z, z, z, z)
\extranewlineforpreprintmode
+ 4 \propV_{\Fbar F\Fbar FS} (z, z, z, 0, z)
+ 4 \propV_{SSSFF}(z, z, z, 0, z)
\nonumber \\ && \quad
+ 4 \propM_{SFS\Fbar F} (z, z, z, z, 0)
+ \propgaugeSS(z,z) + \propgaugeFbarFbar (z,z)/2
]
.
\label{eq:diffRItwo}
\end{eqnarray}
and $\Pi^{(2)}_{I_0 R_0} = 0$.
The right-hand side of equation (\ref{eq:diffRItwo})
vanishes identically when written in terms of the basis functions. 
Therefore, we again obtain a single expression for the self-energy 
function of the neutral complex scalar at two-loop order. 
Written in terms of the basis functions, 
equation (\ref{eq:piIItwo}) 
takes the 
simple form (with $\deltaMSbar = 0$):
\begin{eqnarray}
\Pi^{(2)}_{\phi_0}
&=&
Y^4 \bigl \lbrace
x B(z,z)^2 - 2 (xz+sz+sx) M(z,z,z,z,x)
\extranewlineforpreprintmode
+ (x+s) [2 U(z,z,z,x)-4 z V(z,z,z,x)]
\bigr \rbrace
\nonumber \\ && \!\!
+2 g^2 Y^2
(x+s)
\bigl \lbrace
(2z-s) M(z,z,z,z,0) + 2 \Tbar (0,z,z)
\extranewlineforpreprintmode
+ 4 [A(z)-z] B(z',z)
- 4 B(z,z) -2 \bigr \rbrace .
\label{eq:pineutraltwo}
\end{eqnarray}

Let us now consider the self-energy function of the charged scalar
fields. At one loop order, the diagonal elements are all the same:
\begin{eqnarray}
\Pi_{R_+ R_+}^{(1)} =
\Pi_{I_+ I_+}^{(1)} =
\Pi_{R_- R_-}^{(1)} =
\Pi_{I_- I_-}^{(1)} &=&
Y^2 [ \propA_S(x) + \propA_S(z) +(x+2z) \propB_{SS} (x,z) 
+ \propB_{FF} (x,z)]
\nonumber \\ &&
+ g^2 [\propA_S (z) + 2 \propB_{FF} (0,z) + \propB_{SV}(z,0)]
\end{eqnarray}
while the off-diagonal elements vanish, thanks to supersymmetry:
\begin{eqnarray}
\Pi_{R_+ R_-}^{(1)} = -\Pi_{I_+ I_-}^{(1)} =
Y^2 \sqrt{xz} [ 2 \propB_{SS}(x,z) + \propB_{\Fbar\Fbar}(x,z)] = 0 .
\end{eqnarray}
It follows that the charged scalars have a common self-energy function,
\begin{eqnarray}
\Pi_{\phi_\pm}^{(1)} = 
-Y^2 (s+z) B(x,z) + g^2 [4 z B(0,z) + (1-\xi) \lbrace
(s+z) B(0,z) + 2 A(z) -2s \rbrace] .
\label{eq:pichargedone}
\end{eqnarray}
(The term proportional to $1-\xi$ 
vanishes in the limit 
$s\rightarrow z$.)
The two-loop corrections are, from sections IV and V:
\begin{eqnarray}
\Pi_{R_+ R_+}^{(2)} &=&
Y^4 [
\propS_{SSS}(x, x, z)
+ \propS_{SSS}(z, z, z)
+ \propX_{SSS}(z, z, x)
+ \propX_{SSS}(z, z, z)
+ 2 \propX_{SSS}(x, x, z)
\nonumber \\ && \quad
+ (x + 2 z) \lbrace
\propW_{SSSS}(x, x, z, z)
+ \propW_{SSSS}(z, z, x, z)
\extranewlineforpreprintmode
+ \propY_{SSSS}(x, z, z, x)
+ \propY_{SSSS}(x, z, z, z)
\nonumber \\ && \quad
+ 2 \propY_{SSSS}(z, x, x, z)
+\propZ_{SSSS}(x, z, x, z)
\extranewlineforpreprintmode
+ 2 \propU_{SSSS}(x, z, x, z)
+ 2 \propU_{SSSS}(z, x, z, z)
\nonumber \\ && \quad
+ \propV_{SSSFF}(x, z, z, x, z)
+ \propV_{SSSFF}(z, x, x, z, z)
\rbrace
\extranewlineforpreprintmode
+(x^2 + 4 x z + 8 z^2) \propV_{SSSSS}(z, x, x, z, z)
\nonumber \\ && \quad
+(x^2 + 8 x z + 4 z^2) \propV_{SSSSS}(x, z, z, x, z)
\extranewlineforpreprintmode
+ (8 z x + 4 z^2) \propM_{SSSSS}(x, z, z, x, z)
+ \propW_{SSFF}(x, x, z, z)
\nonumber \\ && \quad
+ \propW_{SSFF}(z, z, x, z)
+z \propV_{F\Fbar \Fbar FS}(x, z, z, x, z)
\extranewlineforpreprintmode
+z \propV_{F\Fbar \Fbar FS}(x, z, z, z, x)
+2 x \propV_{F\Fbar \Fbar FS}(z, x, x, z, z)
\nonumber \\ && \quad
+ \propV_{FFFFS}(x, z, z, x, z)
+ \propV_{FFFFS}(x, z, z, z, x)
\extranewlineforpreprintmode
+2 \propV_{FFFFS}(z, x, x, z, z)
+2 z x \propV_{SSS\Fbar \Fbar }(x, z, z, x, z)
\nonumber \\ && \quad
+2 z^2 \propV_{SSS\Fbar \Fbar }(z, x, x, z, z)
+z \propM_{F\Fbar \Fbar FS}(x, z, z, x, z)
+x \propM_{F\Fbar \Fbar FS}(z, x, x, z, z)
\nonumber \\ && \quad
+2 x \propM_{SFS\Fbar F}(x, z, z, x, z)
+2 z \propM_{SFS\Fbar F}(z, x, x, z, z)
+2 z \propM_{SFSF\Fbar}(x, z, z, x, z)
]
\nonumber \\ &&
+ g^2 Y^2 [
\propX_{SSS}(z, z, x)
+ 2 \propX_{SSS}(z, z, z)
-2 \propS_{SSS}(z, z, z)
+ (4 z-2x) \propU_{SSSS}(z, x, z, z)
\nonumber \\ && \quad
+ (x+2z) \lbrace \propW_{SSSS}(z, z, x, z)
+ \propY_{SSSS}(x, z, z, z)
\extranewlineforpreprintmode
+ 2 \propV_{SSSFF}(x, z, z, 0, z)
+ \propgaugeSSSS (z,z,x,z) 
\rbrace
\nonumber \\ && \quad
+ 2 \propW_{SSFF}(z, z, 0, z)
+ \propW_{SSFF}(z, z, x, z)
\extranewlineforpreprintmode
+2 z \propV_{F\Fbar \Fbar FS}(0, z, z, x, z)
+ 2 z \propV_{F\Fbar \Fbar FS}(0, z, z, z, x)
\nonumber \\ && \quad
+2 z \propV_{F\Fbar \Fbar FS}(x, z, z, 0, z)
+ 2 \propV_{FFFFS}(0, z, z, x, z)
\extranewlineforpreprintmode
+2 \propV_{FFFFS}(0, z, z, z, x)
+ 2 \propV_{FFFFS}(x, z, z, 0, z)
\nonumber \\ && \quad
- 4 \propM_{FFFFS}(0, z, z, x, z)
+ 4 z \propM_{SFS\Fbar F}(z, 0, x, z, z)
\extranewlineforpreprintmode
+ 4 z \propM_{SFSF\Fbar }(x, z, z, 0, z)
+ \propY_{VSSS}(0, z, z, x)
\nonumber \\ && \quad
+ \propY_{VSSS}(0, z, z, z)
+ \propW_{SSSV}(z, z, z, 0)
+ \propgaugeSSFF (z,z,x,z)
]
\nonumber \\ &&
+ g^4 [
\propX_{SSS}(z, z, z)
+ 3 \propS_{SSS}(z, z, z)
+ 2 \propW_{SSFF}(z, z, 0, z)
\extranewlineforpreprintmode
+ 4 \propV_{FFFFS}(0, z, z, 0, z)
+ 8 \propV_{FFFFS}(z, 0, 0, z, z)
\nonumber \\ && \quad
+ 4 z \propM_{F\Fbar \Fbar FS}(0, z, z, 0, z)
+4 z \propV_{F\Fbar \Fbar FS}(0, z, z, 0, z)
\extranewlineforpreprintmode
+ \propW_{SSSV}(z, z, z, 0)
+ \propY_{VSSS}(0, z,z, z)
\nonumber \\ && \quad
+ 2 \propgaugeSSFF (z,z,0,z)
+ \ldots
]
,
\label{eq:pitwoRplusRplus}
\end{eqnarray}
with 
$\Pi_{R_- R_-}^{(2)}$ and
$\Pi_{I_+ I_+}^{(2)}$ and
$\Pi_{I_- I_-}^{(2)}$ the same. 
(The ellipses in equation (\ref{eq:pitwoRplusRplus}) 
indicate that the contribution of
order $g^4$ is not complete, since in this paper I have not
included the contributions with two vector lines.
Terms of order $g^4$ are therefore consistently dropped from here 
on.)
Also,
\begin{eqnarray}
\Pi_{R_+ R_-}^{(2)} &=&
-\Pi_{I_+ I_-}^{(2)} \,=\,
2 Y^4 \sqrt{x z} [
\propW_{SSSS}(z, z, x, z)
+2 \propU_{SSSS}(x, z, x, z)
\extranewlineforpreprintmode
+ 2 \propU_{SSSS}(z, x, z, z)
+ \propY_{SSSS}(x, z, z, x)
\nonumber \\ && \quad
+ \propY_{SSSS}(x, z, z, z)
+2 \propY_{SSSS}(z, x, x, z)
+ \propZ_{SSSS}(x, z, x, z)
\extranewlineforpreprintmode
+ (x+4z) \lbrace \propV_{SSSSS}(x, z, z, x, z)
\nonumber \\ && \quad
+ \propV_{SSSSS}(z, x, x, z, z)
+ \propM_{SSSSS}(x, z, z, x, z) \rbrace
\extranewlineforpreprintmode
+ \propW_{SS\Fbar \Fbar }(z, z, x, z)/2
+ \propV_{\Fbar F\Fbar FS}(x, z, z, x, z)
\nonumber \\ && \quad
+ \propV_{\Fbar F\Fbar FS}(x, z, z, z, x)
+ 2 \propV_{\Fbar F\Fbar FS}(z, x, x, z, z)
\extranewlineforpreprintmode
+\propV_{SSSFF}(x, z, z, x, z)
+\propV_{SSSFF}(z, x, x, z, z)
\nonumber \\ && \quad
+z \propV_{SSS\Fbar \Fbar }(x, z, z, x, z)
+z \propV_{SSS\Fbar \Fbar }(z, x, x, z, z)
\extranewlineforpreprintmode
+ \propM_{FF\Fbar \Fbar S}(x, z, z, x, z)
+ \propM_{SFS\Fbar F}(x, z, z, x, z)
\nonumber \\ && \quad
+ \propM_{SFS\Fbar F}(z, x, x, z, z)
+z \propM_{S\Fbar S\Fbar \Fbar }(x, z, z, x, z)
]
\nonumber \\ &&
+
g^2 Y^2 \sqrt{x z} [
2 \propY_{SSSS}(x, z, z, z)
- 2 \propW_{SSSS}(z, z, x, z)
\extranewlineforpreprintmode
-\propW_{SS\Fbar \Fbar }(z, z, x, z)
+ 4 \propV_{SSSFF}(x, z, z, 0, z)
\nonumber \\ && \quad
- 4 \propM_{F\Fbar F\Fbar S}(0, z, z, x, z)
+4 \propM_{SFSF\Fbar }(x, z, z, 0, z)
+4 \propV_{\Fbar F\Fbar FS}(x, z, z, 0, z)
\nonumber \\ && \quad
+ 2 \propgaugeSSSS (z,z,x,z)
+ \propgaugeSSFbarFbar (z,z,x,z)
]
.
\label{eq:chiracisaweasel}
\end{eqnarray}
Writing these expressions in terms of the two-loop basis functions
with $\deltaMSbar=0$,
we discover that equation (\ref{eq:chiracisaweasel}) 
vanishes identically, 
and equation (\ref{eq:pitwoRplusRplus}) becomes:
\begin{eqnarray}
\Pi_{\phi_\pm}^{(2)} &=&
Y^4 \Bigl \lbrace
z B(x, z)^2
+ (s + z) [
U(x, z, x, z) 
- 2 z V(x, z, x, z) 
+ U(z, x, z, z) 
-2 x V(z, x, z, z) ]
\nonumber \\ &&
- (zx + z^2 + sx + 3 s z) M(x, z, z, x, z)
\Bigr \rbrace
\extranewlineforpreprintmode
+
g^2 Y^2 
\Bigl \lbrace
(8 z^2 - 6 x z -2 s x + 8 s z) M(0, z, z, x, z)
+ 8 z^2 V(0,z,x,z) 
\nonumber \\ &&
-4z U(0,z,x,z)
+(6z+2s) [U(z, 0, z, z) - B(0,z) B(x,z)]
\extranewlineforpreprintmode
+ 2 (s+z) [\Tbar (0, x, z) -2 B(x,z) 
\nonumber \\ &&
+2 \lbrace A(z)-z \rbrace B(x,z')
-1]
+ (1-\xi) \Delta
\Bigr \rbrace
+ {\cal O}(g^4)
\label{eq:pitwoz}
\end{eqnarray}
where
\begin{eqnarray}
\Delta &=&
2 z (s-3z) V(0,z,x,z) + (s+z) \Tbar (0,x,z) + 4 S(0,x,z)
\extranewlineforpreprintmode
-4 z I(x,z,z') - 2 I(x,z,z) - 8 z^2 B(0,z') B(x,z)
\nonumber \\ &&
+ (2s+6z) B(0,z) B(x,z)
+4 [A(z) - 2 z -s] B(x,z) 
\extranewlineforpreprintmode
-2 A(x) - 2 A(z) +2x+2z-s/2
\nonumber \\ &&
+ \bigl \lbrace
(4 z^2 + x z + 4 z s - s x) U(0,z,x,z)
+ 2(x^2 - 8 z^2) T(x,0,z) + 2 z (x-8z) T(z,0,x)
\nonumber \\ &&
+ 8 z [(x-2z) A(x) - x A(z) -xz] B(0,z)/x
-8z^2 \bigr \rbrace/(x-4z) 
\label{eq:defDelta}
\end{eqnarray}
embodies the difference between the Landau gauge and Feynman gauge 
results.
(It should be remarked that $\Delta$ can be rewritten
in various ways that look quite different, using the 
expressions for $B(0,z')$ and $V(0,z,x,z)$ in terms of the basis 
functions;
see equations (3.1) and (3.22) of \cite{Martin:2003qz}.)
The function $\Delta$ can actually be evaluated entirely analytically, 
using the 
expressions
given in section VI of \cite{Martin:2003qz}. The limiting expression for
small $s-z$ 
is particularly simple and useful:
\begin{eqnarray}
\Delta = -2 z B(x,z) [2+2 \lnbar(z-s) -\lnbar z] 
+ \ldots .
\label{eq:approxDelta}
\end{eqnarray}
Here, the ellipses represent terms that vanish as $s \rightarrow z$, 
which can be consistently neglected
since they make only a three-loop order contribution to the self-energy 
function.

The pole squared masses $x_p$ and $z_p$ of the neutral and charged scalar 
fields 
can now be obtained by 
\begin{eqnarray}
x_p &=& x + \widetilde \Pi_{\phi_0} \equiv
x 
+ \frac{1}{16\pi^2} \Pi^{(1)}_{\phi_0} (x) 
+ \frac{1}{(16\pi^2)^2} \left [
\Pi^{(2)}_{\phi_0} (x) +
\Pi^{(1)}_{\phi_0} (x) \Pi^{(1)\prime}_{\phi_0} (x)
\right ]
,
\label{eq:neutralpole}
\\
z_p &=& z + \widetilde \Pi_{\phi_\pm} \equiv
x
+ \frac{1}{16\pi^2} \Pi^{(1)}_{\phi_\pm} (z)
+ \frac{1}{(16\pi^2)^2} \left [
\Pi^{(2)}_{\phi_\pm} (z) +
\Pi^{(1)}_{\phi_\pm} (z) \Pi^{(1)\prime}_{\phi_\pm} (z)
\right ] .
\label{eq:chargedpole}
\end{eqnarray}
[Compare to equations
(\ref{eq:genonelooppole})-(\ref{eq:gentwolooppole}).]
Applying the results above, one finds
\begin{eqnarray}
\Pi^{(1)}_{\phi_0} (x) &=& -2 Y^2 B(z,z) x,
\label{eq:accrefscheat}
\\
\Pi^{(2)}_{\phi_0} (x) + \Pi^{(1)}_{\phi_0} (x) \Pi^{(1)\prime}_{\phi_0}
(x)
& =& Y^4 x \bigl [
-(4z+2x) M(z,z,z,z,x)
+ 4 U(z,z,z,x) 
\extranewlineforpreprintmode
- 8 z V(z,z,z,x)
+ \lbrace 3 B(z,z)
\nonumber \\ &&
- 8 z B(z',z) -4 \rbrace B(z,z)
\bigr ]
+ g^2 Y^2 x \bigl [ (8z-4x) M(z,z,z,z,0) 
\extranewlineforpreprintmode
+ 8 \Tbar (0,z,z)
\nonumber \\ &&
+ 16 \lbrace A(z)-z \rbrace B(z',z)
- 16 B(z,z) -8 \bigr ]
\label{eq:fsucheats}
\end{eqnarray}
with all basis functions on the right evaluated at $s=x$, and
\begin{eqnarray}
\Pi^{(1)}_{\phi_\pm} (z) &=& -2 Y^2 B(x,z) z + 4 g^2 B(0,z) z,
\label{eq:fsuisscum}
\\
\Pi^{(2)}_{\phi_\pm} (z) +
\Pi^{(1)}_{\phi_\pm} (z) \Pi^{(1)\prime}_{\phi_\pm} (z) &=&
Y^4 z \bigl [ -(4z+2x) M(x,z,z,x,z)
+2 U(x,z,x,z) 
\extranewlineforpreprintmode
-4z V(x,z,x,z)
+2 U(z,x,z,z)
\nonumber \\ &&
\!\!\!\!\!\!\!\!\!\!\!\!\!\!\!\!\!\!\!\!\!\!\!\!
\!\!\!\!\!\!\!\!\!\!\!\!\!\!\!\!\!\!\!\!\!\!\!\!
\!\!\!\!
-4x V(z,x,z,z)
+\lbrace 3 B(x,z) -4 x B(x',z) -4 z B(x,z') -4\rbrace B(x,z) \bigr ]
\extranewlineforpreprintmode
+ 8g^2 Y^2 z \bigl [
(2 z-x) M(0,z,z,x,z)
\nonumber \\ &&
\!\!\!\!\!\!\!\!\!\!\!\!\!\!\!\!\!\!\!\!\!\!\!\!
\!\!\!\!\!\!\!\!\!\!\!\!\!\!\!\!\!\!\!\!\!\!\!\!
\!\!\!\!
+ z \lbrace V(0,z,x,z) +B(0,z') B(x,z) \rbrace
+ U(z,0,z,z)
- U(0,z,x,z)/2
+ \Tbar(0,x,z)/2
\nonumber \\ &&
\!\!\!\!\!\!\!\!\!\!\!\!\!\!\!\!\!\!\!\!\!\!\!\!
\!\!\!\!\!\!\!\!\!\!\!\!\!\!\!\!\!\!\!\!\!\!\!\!
\!\!\!\!
+ \lbrace A(z) - z +z B(0,z)\rbrace B(x,z')
\extranewlineforpreprintmode
+ \lbrace 1+ x B(x',z) -3 B(x,z)/2 \rbrace B(0,z) -1/2
\bigr ]  + {\cal O}(g^4)
\label{eq:accrefsarescum}
\end{eqnarray}
with all basis functions evaluated\footnote{As
$s \rightarrow z$, the functions $V(0,z,x,z)$
and $B(0,z') B(x,z)$ each have a logarithmic
divergence, but their sum is well-defined and
can be evaluated analytically using equations in
section VI of \cite{Martin:2003qz}.}
at $s=z$.
These results satisfy several non-trivial consistency checks, as
follows.

First, the mere fact that we can write diagonal self-energy functions
$\Pi_{\phi_0}$ and $\Pi_{\phi_\pm}$ is actually a supersymmetric
consequence of the fact that the real and imaginary components of each of
$\phi_0$, $\phi_+$ and $\phi_-$ reside in supermultiplets.
(In a non-supersymmetric theory, the real and imaginary components of
the uncharged field
$\phi_0$ receive different self-energy corrections, and there
is mixing in the self-energy functions between $\phi_+$ and
$\phi_-^*$.)

Next, consider the fact that the pole squared masses $x_p$ and $z_p$ must
be gauge invariant, since they are physical observables. This requires
\end{widetext}
\begin{eqnarray}
\frac{\partial}{\partial \xi} \widetilde \Pi_{\phi_0} =0;
\qquad\qquad
\frac{\partial}{\partial \xi} \widetilde \Pi_{\phi_\pm} =0.
\label{eq:gichecks}
\end{eqnarray}
These equations are indeed seen to be satisfied, since $\xi$ does not
appear in
equations (\ref{eq:accrefscheat})-(\ref{eq:accrefsarescum}).
Let us examine how this happened.
For the neutral scalars, the one-loop gauge
invariance is trivial, while the two-loop gauge invariance can also be
seen from the fact that equation (\ref{eq:piIItwo}) is written in terms of
the gauge-invariant combinations $\propgaugeSS(z,z)$, $\propgaugeFF(z,z)$
and $\propgaugeFbarFbar(z,z)$ that were defined in equations
(\ref{eq:defpropgaugeSS}), (\ref{eq:defpropgaugeFF}) and
(\ref{eq:defpropgaugeFbarFbar}).
To see how the gauge invariance of the charged scalar pole
squared mass $z_p$ comes about, note that using
the analytical expressions for $A(z)$ and $B(0,z)$ in
equation (\ref{eq:pichargedone}) gives
\begin{eqnarray}
\frac{\partial
}{\partial\xi}
\Pi^{(1)}_{\phi_\pm} (s)=
g^2 (1-\frac{z}{s}) [ (z+s) \lnbar(z-s) -z \lnbar z ],
\phantom{iii}
\label{eq:dpipmonedxi}
\end{eqnarray}
from which the limit
\begin{eqnarray}
\frac{\partial}{\partial\xi} \Pi^{(1)}_{\phi_\pm} (z) =0
\end{eqnarray}
follows immediately, and
\begin{eqnarray}
&&
\frac{\partial}{\partial\xi}
\left [ \Pi^{(1)}_{\phi_\pm} (s) \Pi^{(1)\prime}_{\phi_\pm} (s) \right ]
=
\nonumber \\ &&
-2 g^2 Y^2 z B(x,z)\left [ 2+2 \lnbar(z-s)- \lnbar z  \right ]
+ \ldots ,
\phantom{xxxx}
\label{eq:dpipmtwodxi}
\end{eqnarray}
where the ellipses indicate terms that vanish as $s\rightarrow z$
and terms of order $g^4$.
Combining equation (\ref{eq:dpipmtwodxi}) with (\ref{eq:pitwoz}) and
(\ref{eq:approxDelta}) gives the desired  smooth limit as
$s \rightarrow z$:
\begin{eqnarray}
\frac{\partial}{\partial\xi}
\left [
\Pi^{(2)}_{\phi_\pm} (z) + \Pi^{(1)}_{\phi_\pm} (z)
\Pi^{(1)\prime}_{\phi_\pm} (z) \right ]
=0 .
\end{eqnarray}
This cancellation explains the absence of terms proportional to $(1-\xi)$
in equation (\ref{eq:accrefsarescum}).

As another check, suppose that $x=0$; then the mass of the neutral fermion
$\psi_0$
is zero at tree-level, and is protected from corrections by a
chiral symmetry. It follows from supersymmetry that the scalar squared
mass also vanishes, so $x_p=0$. This is checked, since
equations (\ref{eq:accrefscheat}) and (\ref{eq:fsucheats}) each vanish
when
$x=0$.

Similarly, suppose that $z=0$. Then the masses of the charged fermions
vanish, and we must have $z_p =0$ to all orders in perturbation theory.
Again, this checks, since equations
(\ref{eq:fsuisscum}) and (\ref{eq:accrefsarescum}) each vanish when $z=0$.

Another important consistency check on the preceding results is provided
by renormalization group invariance of the pole masses of the scalars.
The beta functions of the parameters of the theory can be written in the
general form:
\begin{eqnarray}
\beta_X \equiv Q \frac{d X}{d Q} =
\frac{1}{16 \pi^2} \beta^{(1)}_X +
\frac{1}{(16 \pi^2)^2} \beta^{(2)}_X + \ldots ,
\label{eq:genbeta}
\end{eqnarray}
where
\begin{eqnarray}
\beta^{(1)}_x &=& 4 Y^2 x ,
\label{eq:betaxone}
\\
\beta^{(2)}_x &=& [16 g^2 Y^2 - 8 Y^4] x ,
\label{eq:betaxtwo}
\\
\beta^{(1)}_z &=& [4 Y^2 - 8 g^2] z ,
\label{eq:betazone}
\\
\beta^{(2)}_z &=& [32 g^4  - 8 Y^4] z ,
\label{eq:betaztwo}
\\
\beta^{(1)}_Y &=& 3 Y^3 - 4 g^2 Y ,
\label{eq:betaYone}
\\
\beta^{(2)}_Y &=& -6 Y^5 + 4 g^2 Y^3 +16 g^4 Y ,
\label{eq:betaYtwo}
\\
\beta^{(1)}_g &=& 2 g^3,
\label{eq:betagone}
\\
\beta^{(2)}_g &=& 8 g^5 - 4 g^3 Y^2 .
\label{eq:betagtwo}
\end{eqnarray}
The requirement that the pole squared masses must not depend on the
renormalization scale $Q$ implies:
\begin{eqnarray}
\beta_x &=& -Q\frac{d}{dQ}\widetilde \Pi_{\phi_0} ,\\
\beta_z &=& -Q\frac{d}{dQ}\widetilde \Pi_{\phi_\pm} .
\label{eq:checkbetax}
\end{eqnarray}
Let us check these. First, for the neutral scalars,
we have from equation (\ref{eq:pineutralone}):
\begin{eqnarray}
Q\frac{d}{dQ} \Pi_{\phi_0}^{(1)}(x) &=& -2 Y^2 x\,
Q\frac{\partial B(z,z)}{\partial Q}
- 4 \beta_Y Y x B(z,z)
\nonumber \\ &&
- 2 Y^2 \beta_x \left [B(z,z) + x
\frac{\partial B(z,z)}{\partial s} \right ]
\nonumber \\ &&
- 4 Y^2 x \beta_z B(z,z') ,
\end{eqnarray}
where all  functions on the right side are to be
evaluated at $s=x$.
Using equations (\ref{eq:dBdQ}), (\ref{eq:betaxone}), (\ref{eq:betazone}),
(\ref{eq:betaYone}) and (\ref{eq:dBds}),
this becomes
\begin{eqnarray}
Q\frac{d}{dQ} \Pi_{\phi_0}^{(1)} (x) &=&
-4 Y^2 x + \frac{1}{16 \pi^2} \Bigl \lbrace
Y^4 \left [ 8 - 20 B(z,z)\right ]
\nonumber \\ &&
\!\!\!\!\!
\!\!\!\!\!
+ 16 g^2 Y^2 [ B(z,z) + 2 z B(z',z) ] \Bigr \rbrace x
\phantom{xxx}
\label{eq:betapineutone}
\end{eqnarray}
up to terms that contribute only at three-loop order.
Meanwhile, applying equations
(\ref{eq:dAdQ})-(\ref{eq:dMdQ}),
we obtain
\begin{eqnarray}
&&Q\frac{d}{dQ} \left [\Pi_{\phi_0}^{(1)} (x) \Pi_{\phi_0}^{(1)\prime }
(x) \right ] =
8 Y^4 \bigl [B(z,z)
\nonumber \\ && \qquad\qquad -2z B(z',z) -1\bigr ] x
,
\phantom{xxx}
\label{eq:betapineutoneone}
\\
&&Q\frac{d}{dQ} \Pi_{\phi_0}^{(2)} (x) =
Y^4 [ 8 + 12 B(z,z) + 16 z B(z',z) ] x
\nonumber \\ &&
\qquad\>
-16 g^2 Y^2 [ 1 + B(z,z) + 2 z B(z',z)] x ,
\phantom{xx}
\label{eq:betapineuttwo}
\end{eqnarray}
again up to terms that contribute only at three-loop order, and with
$s=x$ in the functions on the right side. Combining
equations
(\ref{eq:betapineutone}), (\ref{eq:betapineutoneone}) and
(\ref{eq:betapineuttwo}) gives
\begin{eqnarray}
Q\frac{d}{dQ} \widetilde \Pi_{\phi_0}
&=& \frac{1}{16 \pi^2}[ -4 Y^2 x]
\nonumber \\ &&
\!\!\!\!\!
\!\!\!\!\!
\!\!
+ \frac{1}{(16 \pi^2)^2}[ 8 Y^4 - 16 g^2 Y^2 ] x  + \ldots
.
\end{eqnarray}
This verifies that equation (\ref{eq:checkbetax}) is indeed consistent
with equations (\ref{eq:betaxone}) and (\ref{eq:betaxtwo}).

In the same way, one can check renormalization scale invariance of the
complex scalar pole squared mass. One finds
\begin{eqnarray}
Q\frac{d}{dQ} \Pi_{\phi_\pm}^{(1)} (z) &=&
[8 g^2 - 4 Y^2]z
+
\frac{1}{16 \pi^2} \bigl \lbrace
Y^4 [ 8-20 B(x,z)]
\nonumber \\ &&
\!\!\!\!\!\!\!
\!\!\!\!\!\!\!
\!\!\!\!
\!\!\!\!
+ 16 g^2 Y^2 \left [
2 B(x,z) - x B(x',z)
-\lnbar z
\right ] \bigr \rbrace
z \phantom{xxx}
\end{eqnarray}
and
\begin{eqnarray}
Q\frac{d}{dQ} \left [ \Pi_{\phi_\pm}^{(2)}(z)
+ \Pi_{\phi_\pm}^{(1)} (z) \Pi_{\phi_\pm}^{(1)\prime } (z) \right ] =
20 Y^4 z B(x,z)
&&
\nonumber \\
+ 16 g^2 Y^2 \left [
-2 B(x,z)
+x B(x',z)
+\lnbar z
\right ] z
,
\phantom{xxx}
&&
\label{eq:QdQpitwoz}
\end{eqnarray}
up to terms of order $g^4$ and terms of three-loop order and with $s=z$
in all loop functions on the right side.
From
these two equations, we obtain
\begin{eqnarray}
Q\frac{d}{dQ} \widetilde \Pi_{\phi_\pm} &=&
\frac{1}{16 \pi^2} \bigl [8 g^2 -4 Y^2 ] z
\nonumber \\ &&
+\frac{1}{(16 \pi^2)^2} \bigl [8 Y^4 + {\cal O}(g^4)] z,
\end{eqnarray}
in successful agreement with equations (\ref{eq:betazone}) and
(\ref{eq:betaztwo}).

Finally, for a numerical study, consider the pole squared masses of the
neutral scalars, for a model defined by running parameters $g=Y=1$ and
$x=1$, $z=0.1$ at a renormalization scale $Q_0=1$ (in arbitrary
units). These choices allow the decays
$\phi_0 \rightarrow \phi_+\phi_-$
and
$\phi_0 \rightarrow \psi_+\psi_-$
and
$\phi_0 \rightarrow \phi_+\phi_- A$
and
$\phi_0 \rightarrow \psi_+\psi_- A$
and
$\phi_0 \rightarrow \psi_+\phi_- \chi$,
and
$\phi_0 \rightarrow \psi_-\phi_+ \chi$, so the pole squared mass will
have an
imaginary part corresponding to the total width. The value
of equation (\ref{eq:neutralpole}) with (\ref{eq:accrefscheat}) and
(\ref{eq:fsucheats})
can be found numerically by computing the master integral
$M(z,z,z,z,x)$
simultaneously with its subordinates $U(z,z,z,x)$, $T(z,x,z)$, and
$S(x,z,z)$, [and thus $V(z,z,z,x)$] as described in \cite{Martin:2003qz}.
The function
$M(z,z,z,z,0)$ can be computed analytically (as found first in
\cite{Broadhurst:1987ei}, and presented in the notation of the present
paper in equation (6.27) of
\cite{Martin:2003qz}), as can $\Tbar(0,z,z)$, $I(x,z,z)$, $B(z,z)$,
$B(z',z)$, $A(z)$
and $A(x)$.
I then find, working at $Q=Q_0=1$:
\begin{eqnarray}
\left [M^2 - i \Gamma M \right
]_{\phi_0,\,\text{1-loop}} &=&
%0.96575023 - 0.030820222\, i
0.96575 - 0.03082\, i,
\phantom{xxx}
\\
\left [M^2 - i \Gamma M \right
]_{\phi_0,\, \text{2-loop}} &=&
%0.96423154 - 0.033120256\, i .
0.96423 - 0.03312 \, i  .
\phantom{xxx}
\end{eqnarray}
\begin{figure}[tb]
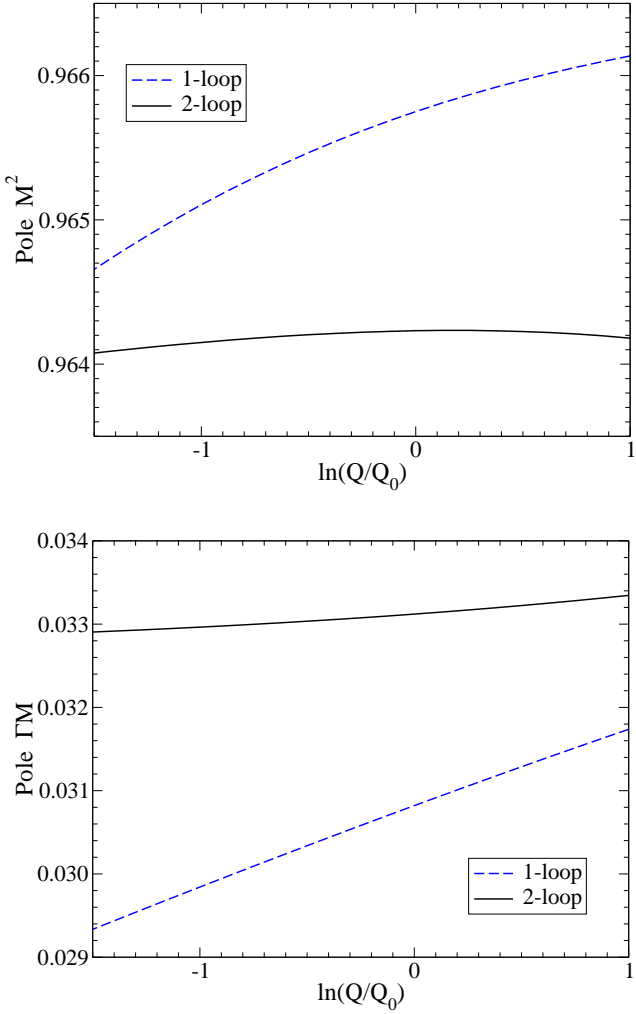

\includegraphics[width=8.4cm]{susyre}
\vspace{0.2in}

\includegraphics[width=8.25cm]{susyim}
\caption{\label{fig:numexample}The real part, $M^2$, and the magnitude
of the imaginary part, $\Gamma M$, of the neutral scalar pole squared mass
are shown as a function of the logarithm of the renormalization scale
$Q$, for the model described in the text.
The dashed lines are the one-loop approximation, and the solid lines are
the two-loop approximation.}
\end{figure}
Now, to test the renormalization group scale-independence that was
formally checked above, one can run the parameters $x,z,Y,g$ from the
scale $Q_0=1$ to any other scale $Q$ using equations
(\ref{eq:genbeta})-(\ref{eq:betagtwo}), and then recompute the complex
pole squared mass in the same way. 
The results are shown in Figure
\ref{fig:numexample}. The two-loop approximation to the pole squared mass
features a significantly better scale dependence than the one-loop
approximation, as one might expect.

\protect\section{\label{sec:outlook}\protect{Outlook}}
\setcounter{equation}{0}

%``Stupidity is reaching a conclusion."

In this paper, I have presented results for the two-loop self-energy
functions for scalars in a general renormalizable theory, including the
contributions of all Feynman diagrams that contain only one vector
propagator (or none). This is equivalently the leading non-trivial order, 
quadratic,
in gauge couplings.  In some cases, notably Higgs scalar boson masses in
the electroweak Standard Model and most extensions of it, these are the
dominant two-loop contributions, because electroweak gauge couplings are
smaller than the QCD and top Yukawa (and possibly bottom Yukawa)
couplings. I have specialized the results obtained here to the Higgs
scalar bosons of the Minimal Supersymmetric Standard Model, and the
results will appear elsewhere \cite{toappear}.

Further progress will require evaluating the contributions of the
remaining two-loop Feynman diagrams, containing two or more vector
propagators. A hallmark of the strategy used here is that it is flexible,
applying to a general renormalizable field theory. In softly broken
supersymmetry, the full two-loop result should allow, as special cases,
precise evaluation of the pole masses of not only the Higgs scalar bosons,
but also the squarks and sleptons.

\protect\section*{\label{sec:appendix}\protect{Appendix}}
\renewcommand{\theequation}{A.\arabic{equation}}
\setcounter{equation}{0}

This Appendix contains a few useful identities involving the basis
functions. Reference \cite{Martin:2003qz} contains many more identities,
including analytic expressions for some of the basis functions, and the
derivatives of all of the basis functions with respect to $s$ and each
squared-mass argument, expressed algebraically in terms of the basis
functions.

Partial derivatives with respect to the renormalization scale are given
by:
\begin{eqnarray}
Q \frac{\partial}{\partial Q} A(x) &=& -2x
\label{eq:dAdQ}
,
\\
Q \frac{\partial}{\partial Q} B(x,y) &=& 2
,
\label{eq:dBdQ}
\\
Q \frac{\partial}{\partial Q} I(x,y,z) &=&
2[A(x)+A(y)+A(z)
\nonumber \\ &&
-x-y-z]
,
\label{eq:dIdQ}
\\
Q \frac{\partial}{\partial Q} S(x,y,z) &=&
2[A(x)+A(y)+A(z)
\nonumber \\ &&
-x-y-z] +s
,
\label{eq:dSdQ}
\\
Q \frac{\partial}{\partial Q} T(x,y,z) &=& -2A(x)/x
,
\label{eq:dTdQ}
\\
Q \frac{\partial}{\partial Q} \Tbar (0,x,y) &=& 2 - 2B(x,y)
,
\label{eq:dTbardQ}
\\
Q \frac{\partial}{\partial Q} U(x,y,z,u) &=&
2+2B(x,y)
,
\label{eq:dUdQ}
\\
Q \frac{\partial}{\partial Q} V(x,y,z,u) &=&
-2B(x,y')
,
\phantom{\Biggl [ \Biggr ]}
\label{eq:dVdQ}
\\
Q \frac{\partial}{\partial Q} \Vbar (x,0,z,u) &=&
\lbrace 2 (s+x) [B(0,x)-1]
\nonumber \\ &&
+ 4 A(x) \rbrace/(s-x)^2
,
\label{eq:dVbardQ}
\\
Q \frac{\partial}{\partial Q} M(x,y,z,u,v) &=& 0
.
\label{eq:dMdQ}
\end{eqnarray}

The function $\Vbar(x,0,u,z)$ introduced in equation (\ref{eq:defVbar})
can be rewritten explicitly in terms of the basis integrals as:
\begin{widetext}
\begin{eqnarray}
&&\Vbar (x,0,z,u) =
-x U(x,0,z,u)/(s-x)^2 
+\left [ \frac{z+u}{(z-u)^2(s-x)} - \frac{1}{(s-x)^2} \right ] x T(x,z,u)
\nonumber \\ && \qquad
+\left [ \frac{u(z-u-x+s)}{(z-u)^3 (s-x)} -\frac{2x}{(z-u)(s-x)^2} 
  \right ] z T(z,x,u)
\extranewlineforpreprintmode
+\left [ \frac{z(z-u+x-s)}{(z-u)^3 (s-x)} +\frac{2x}{(z-u)(s-x)^2} 
  \right ] u T(u,x,z)
\nonumber \\ && \qquad
+ 
\Bigl [
(z+u) \lbrace S(x,z,u) -A(x)/2 +x+z+u-5s/8 \rbrace + A(z) A(u) \Bigr ]
/[(z-u)^2(s-x)]
\nonumber \\ && \qquad 
+ B(0,x) \left [
\frac{(2x+s) \lbrace A(u)-A(z) \rbrace}{(s-x)^2 (z-u)} 
+ \frac{u A(z) - z A(u)}{(z-u)^3}
- \frac{(z+u)}{2(z-u)^2} \right ]
\extranewlineforpreprintmode
+ \frac{A(x) \lbrace A(u) - A(z) \rbrace}{(z-u)(s-x)^2}
\nonumber \\ && \qquad 
+ 
\left [
(xz+xu-2su) A(z) 
+ (xz+xu-2sz) A(u) 
\right ]/[(z-u)^2(s-x)^2]
,
\label{eq:defVbarxozu}
\\
&&
\Vbar (x,0,z,z) =
\frac{1}{3 z (s-x)^2} \Bigl [
(4z-3x-s) z T(z,z,x)
+(s-x) x T(x,z,z) 
\extranewlineforpreprintmode
+ (4z-x+s) S(x,z,z)
-2 z I(0,z,z)
\nonumber \\ && \qquad
+ 2 A(z) \lbrace x-2z-A(x)\rbrace 
+ A(x) \lbrace x/2-4z-s/2 \rbrace
+ 4 z^2-z x - x^2 
\extranewlineforpreprintmode
+ 13 s x/8 + s z/2 - 5 s^2/8
\Bigr ]
\nonumber \\ && \qquad
- B(0,x) \left [
\lbrace A(z)/z+1\rbrace (s+x)/(s-x)^2 + 1/6z \right ]
,
\\ &&
\Vbar (x,0,0,0) = \Bigl [ -(s+x) \Tbar(0,0,x) + 2 S(0,0,x)
-(s+x) B(0,x) -3 A(x) 
\extranewlineforpreprintmode
+x-s/4 \Bigr]/(s-x)^2 .
\end{eqnarray}
These functions are singular as $s \rightarrow x$, but the functions
$\propV_{FFFFS}(x,0,0,u,v)$ in 
eq.~(\ref{VFFFFSxoouv})
and 
$\propV_{FFFFV}(x,0,0,u,v)$ in 
eq.~(\ref{VFFFFVxoouv})
are smooth in that limit.
\vspace{0.25in}

Some identities involving vanishing squared-mass arguments are:
\begin{eqnarray}
U(x,y,y,0) &=& - T(y,0,x) + [1 - A(y)/y] B(x,y) +1
,
\label{eq:UxyyO}
\\
U(x,0,0,0) &=& -\Tbar(0,0,x) + 2 B(0,x) +1
,
\\
V(x,y,y,0) &=& \bigl \lbrace 
\Tbar (0,x,y)  - T (y,0,x) - [A(y)/y+1] B(x,y)
\extranewlineforpreprintmode
   + 2 [A(y)-y] B(x,y') \bigr \rbrace/2y
,
\\
U(x,0,y,z) &=& 
\left[ 1/(z-y) + 1/(s-x) \right ] y T(y,x,z) 
+\left[ 1/(y-z) + 1/(s-x) \right ] z T(z,x,y)
\nonumber \\ &&
+ \bigl [ 2 x T(x,y,z) + 2 S(x,y,z) - I(0,y,z) -A(x)
\extranewlineforpreprintmode
-A(y)-A(z)
+x+y+z-s/4 \bigr ]/(s-x)
\nonumber \\ &&
+ B(0,x) [A(y)-A(z)]/(z-y) 
,
\\
U(x,0,y,y) &=& \bigl [
(s-x-4y) T(y,y,x) - 4 x T(x,y,y) - 4 S(x,y,y) 
\extranewlineforpreprintmode
+ 2 I(0,y,y) + 2 A(x)
+ 4 A(y)
\nonumber \\ &&
-3x-4y+3s/2 \bigr ]/(x-s) - [1 + A(y)/y]B(0,x)
\\
4 S(0,x,y) &=&
(x-3y-s) T(y,x,0) + (y-3x-s) T(x,y,0) 
\extranewlineforpreprintmode
+ I(0,y,y)
+ I(0,x,x) - s B(x,y)^2
\nonumber \\ &&
\!\!\!\!\!
\!\!\!\!\!
\!\!\!\!\!
+ B(x,y) [ 2s + (y-x-s) A(x)/x +(x-y-s) A(y)/y]
\extranewlineforpreprintmode
+ 2 [A(x) + A(y) -x-y)] + 3 s/2 
, \phantom{xxx}
\\ 
S(0,0,x) &=&
-x T(x,0,0) + (s-x) B(0,x)/2 + A(x)/2-x + 5s/8
.
\label{eq:SOOx}
\end{eqnarray}
\end{widetext}
The functions in the preceding identities appear in the
self-energy functions involving massless vector propagators, making the
presentation of the formulas given in section \ref{sec:massless} quite
non-unique.

This work was supported by the National Science Foundation under Grant No.
0140129.

\end{document}